\documentclass[prd,aps]{revtex4}
\usepackage{graphicx,latexsym}
\usepackage{epsfig}
\usepackage{amssymb}
\usepackage{amsfonts}
\usepackage{amsmath}
\usepackage{euscript}
\usepackage{verbatim}
\usepackage{latexsym}
\usepackage{graphicx}
\usepackage{caption}
\usepackage{float}
\usepackage{wrapfig}
	\usepackage[T1]{fontenc}
	\usepackage{tikz}
	\usetikzlibrary{decorations.pathmorphing}
\usepackage{tikz}
\usetikzlibrary{shapes.geometric, arrows,patterns,snakes}
\tikzstyle{ellip} = [ellipse, minimum width=3cm, minimum height=1cm,text centered, draw=black]

\newcommand{\beq}{\begin{equation}}
\newcommand{\eeq}{\end{equation}}
\newcommand{\bea}{\begin{eqnarray}}
\newcommand{\eea}{\end{eqnarray}}
\newcommand{\beas}{\begin{eqnarray*}}
\newcommand{\eeas}{\end{eqnarray*}}

\newcommand{\non}{\nonumber}
\newcommand{\bquo}{\begin{quote}}
\newcommand{\enqu}{\end{quote}}
\renewcommand{\(}{\begin{equation}}
\renewcommand{\)}{\end{equation}}
\def\diag{\hbox{\rm diag}}

\begin{document}

\title{(Anti-)Symmetrizing Wave Functions} 

\author{Chethan Krishnan$^{1}$, K. V. Pavan Kumar$^{2}$,
and P. N. Bala Subramanian$^{3}$} 

\affiliation { Center for High Energy Physics,\\
Indian Institute of Science, Bangalore 560012, India}

\begin{abstract}
The construction of fully (anti-)symmetric states with many particles, when the single particle state carries multiple quantum numbers, is a problem that seems to have not been systematically addressed in the literature. A quintessential example is the construction of ground state baryon wave functions where the color singlet condition reduces the problem to just two (flavor and spin) quantum numbers. In this paper, we address the general problem by noting that it can be re-interpreted as an eigenvalue equation, and provide a formalism that applies to generic number of particles and generic number of quantum numbers. As an immediate result, we find a complete solution to the two quantum number case, from which the baryon wave function problem with arbitrary number of flavors follows. As a more elaborate illustration that reveals complications not visible in the two quantum number case, we present the complete class of states possible for a system of five fermionic particles with three quantum numbers each. Our formalism makes systematic use of properties of the symmetric group and Young tableaux. Even though our motivations to consider this question have their roots in SYK-like tensor models and holography, the problem and its solution should have broader applications. 
\end{abstract}

\maketitle
\section{The Problem}

The goal of this paper is to provide a systematic procedure for constructing all possible states made of a fixed number of fermionic\footnote{We will mostly deal with fermions for concreteness, but an entirely analogous discussion holds for bosons as well.} tensors of the form $\psi ^{i\ldots j}$. The various indices on these tensors can be thought of as proxies for various quantum numbers these fermions carry, and the number of tensors can be viewed as the number of particles. Due to the fermionic nature of $\psi $'s, only the representations that are antisymmetric under the exchange of any two fermions will arise. In other words, the problem we wish to solve is closely related to the question of finding the fully anti-symmetric multi-particle representations of the group $G_i\times \ldots \times G_j$ where each index ($i$) of the tensor transforms in the fundamental of the corresponding group ($G_i$). We will think of the groups $G_i$ as $U(N)$ with possibly distinct $N$'s in each slot, but our strategy should be adaptable to arbitrary groups with minor modifications.  

We will explicitly find multi-particle states for fermions carrying two and three quantum numbers as an illustration of our approach.  The two quantum number case has some extra simplifications. As a more elaborate illustration of our technology, we will also present explicit results for the cases with four and five particles, each carrying three quantum numbers. It is conceptually straightforward, but possibly computationally challenging\footnote{Calculations involving Young tableaux are claimed \cite{Fuchs} to be non-trivial even on a computer.} to extend it to higher number of particles and quantum numbers per particle. But we will formulate the problem as an eigenvalue problem, so we emphasize that in principle  it is tractable in full generality. However, our aim in the later sections will be to find nice results at low levels and small number of quantum numbers. 

Throughout this paper, we will mostly deal with fermionic systems for concreteness\footnote{See also our discussion in the next paragraph for our motivations for considering this problem.}, but we will present one bosonic case. This will be the bosonic case with two quantum numbers, and it is closely related to the problem of the construction of ground state wave functions for baryons: ground state means that we take the two independent orbital angular momenta in the 3-quark system to be vanishing ($\ell=\ell'=0$). This is a problem well-known from introductory particle physics courses, but let us quickly review it here for completeness. The relevant quantum numbers in the $\ell=\ell'=0$ state are color, flavor and spin, and because we expect baryons to be color singlets, the problem effectively reduces to a two quantum number problem. Since an $SU(3)$ color singlet made from three fundamentals is fully anti-symmetric, the problem reduces to the construction of states with two quantum numbers (flavor and spin) that are {\em symmetric} under the interchange of any two particles. Therefore the {\em bosonic} two quantum number case that we will write down using our approach subsumes the solution to the baryon wave function problem\footnote{For three flavors, the baryon wave function problem is discussed in many introductory books on particle physics, eg. \cite{Griffiths}. But we have not a found a discussion in the literature which includes more flavors.}.

Our motivation for considering this problem arose from investigations of  certain classes of quantum mechanical tensor models where the symmetry group above arises as a global or gauged symmetry \cite{Witten, Klebanov}. It was noticed in \cite{our1, our2, finiteN, our4} that for low values of the rank of the group, these models can potentially be solved at least on a computer. Our discussions in this paper are directly relevant to solving the ungauged models following the approach of \cite{finiteN}, but we will not further discuss this application in this paper and merely restrict our attention to the mathematical problem. See also some discussions in the gauged theory, which use loosely similar group theory techniques \cite{MelloKoch,Ramgoolam}. 

Let us consider the 3-index fermions of the form $\psi ^{ijk}$. The indices $\{i,j,k\}$ can be taken to belong to the group $SU(n_1)_i\times SU(n_2)_j\times SU(n_3)_k$ and therefore take values from 1 to $n_{1,2,3}$. More precisely, $\psi ^{ijk}$  transform under the vector representation of each of $SU(n_i)$ i.e.,
\begin{align}
\psi ^{ijk}\rightarrow M_1 ^{ii'}M_2^{jj'}M_3^{kk'} \psi ^{i'j'k'}
\end{align}     
where $M_1,M_2,M_3$ belong to the three $SU(n_i)$'s respectively. A general state involving $n$ fermions is of the form:
\begin{align}
\psi ^{i_1j_1k_1}\psi ^{i_2j_2k_2}\ldots \psi ^{i_nj_nk_n}
\end{align}
This state is antisymmetric under exchange of any two fermions i.e., 
\begin{align}
\psi ^{i_1j_1k_1}\ldots \psi ^{i_aj_ak_a}\ldots \psi ^{i_nj_nk_n}=-\psi ^{i_aj_ak_a}\ldots \psi ^{i_1j_1k_1}\ldots \psi ^{i_nj_nk_n}
\end{align}
The states can be organized in terms of irreducible representations of $SU(n_1)_i\times SU(n_2)_j\times SU(n_3)_k$. Because of the fermionic nature, some of the representations become trivially zero. Our goal is to find a systematic way to find all the non-trivial representations that the fermionic states fall into.

This question is most easily answered in terms of Young tableaux. In the language of Young tableaux, a general state at level $n$ can be written as:
\begin{figure}[h]
	\centering
	\includegraphics[trim={2cm 26cm 3.7cm 2.5cm},clip,scale=0.9]{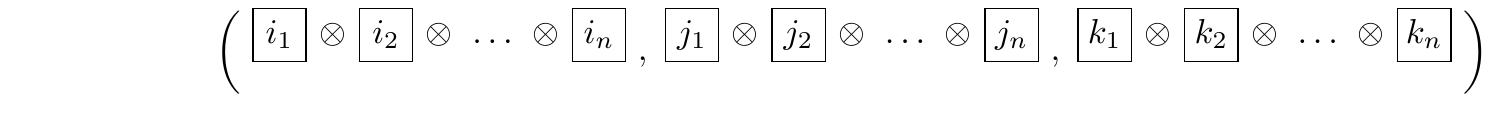}
\end{figure}
%\begin{align}
%\left(~\begin{ytableau}
%i_1 &\none[\otimes] &i_2 &\none[\otimes]&\none[\dots]&\none[\otimes] &i_n
%\end{ytableau}~ ,
%~\begin{ytableau}
%j_1 &\none[\otimes] &j_2 &\none[\otimes]&\none[\dots]&\none[\otimes] &j_n
%\end{ytableau}~,~ \begin{ytableau}
%k_1 &\none[\otimes] &k_2 &\none[\otimes]&\none[\dots]&\none[\otimes] &k_n
%\end{ytableau}~ \right)
%\end{align} 
\\
The number of quantum numbers becomes the number of {\em slots}, and the number of particles (which we will sometimes call the {\em level}) becomes the number of boxes in each slot. The representation content of each slot can be figured out by decomposing the tensor products into various irreducible representations via Littlewood-Richardson rules. The question we want to answer is what are the irreps that survive in the full object after we impose anti-symmetry under exchange of particles. 

We will answer this question by working with permutation groups $S^{(i)}_n\times S^{(j)}_n\times S^{(k)}_n$ (where $S_n$ stands for the permutation group with $n$ elements) instead of $G_i\times G_j\times G_k$. %More precisely, if we assume that none of the indices corresponding to any two different fermions are same then at level $r$, we can work with the irreps of $S^{(i)}_r\times S^{(j)}_r\times S^{(k)}_r$. 
If we wish to work with a specific group, we can impose further constraints on the allowed representations (aka. Young tableaux) that show up. %In principle, this means that the solution with some indices being equal is a subset of the solution we present here. 
Lets illustrate this with a simple example: let us consider as case where we are working with $U(3)$ groups, but looking at level $ \ge 4$. In this case, because there are not enough indices to soak up all the slots in the tableaux, (for example) some of the representations will be zero. So the general problem we solve, together with the specific restrictions on Young tableaux that arise for the specific group will be the complete solution of our problem for that group. In the final section, we will use an argument based on the group $SU(n)$, as a useful sanity check of our results. We have collected some useful facts about the symmetric group and its representations in an Appendix.

Let $R_i$, $R_j$ and $R_k$ denote the irreducible representations (as can be captured by Young patterns) of the corresponding permutation groups. Then their tensor product $R_i\times R_j \times R_k$ are irreducible representations under $S^{(i)}_n\times S^{(j)}_n\times S^{(k)}_n$. We need the irreps of 
$S^{(i)}_n\times S^{(j)}_n\times S^{(k)}_n$ such that they are antisymmetric under exchange of any two objects. More operationally, the required irreps need to satisfy the following equation\footnote{Lets emphasize once again that for most of the paper we will stick to the negative sign on the right hand side, which corresponds to fermions.}: 
\begin{align}
\label{main equation}
D(g)_{R_i}\otimes D(g)_{R_j}\otimes D(g)_{R_k}&\left[\sum_{i,j,k} \alpha _{ijk}^{(R_i,R_j,R_k)} ~|i\rangle _{R_i}\otimes |j\rangle _{R_j}\otimes |k\rangle _{R_k}\right]\non\\
&\qquad =\pm\sum_{i,j,k} \alpha _{ijk}^{(R_i,R_j,R_k)} ~|i\rangle _{R_i}\otimes |j\rangle _{R_j}\otimes |k\rangle _{R_k}
\end{align}
This is our main equation, and by writing this equation, we have translated our problem into an eigen-problem. Here $g$ is one of the transpositions\footnote{We note that the any other 2-cycle (and therefore all elements of the group) can be obtained from $(12),(23),\ldots (n-1,n)$} (2-cycles) of the form $(i,i+1)$ for $i=1,\ldots (n-1)$.  $D(g)_{R_i}$, $D(g)_{R_j}$ and $D(g)_{R_k}$ are the matrix forms of $g$ in the representations $R_i$, $R_j$ and $R_k$ respectively. $|i\rangle _{R_i}$, $|j\rangle _{R_j}$ and $|k\rangle _{R_k}$ denote the standard Young tableaux of the representations $R_i$, $R_j$ and $R_k$ respectively and the summation is taken over all the standard Young tableaux. We have written the equation for the three slot/index case, but it should be clear that this equation straightforwardly generalizes to more indices.

The claim is that solving the above equation for $\alpha$'s will accomplish the solution to the problem we stated in the beginning of this section. Note that once formulated in this manner in the language of symmetric groups and its representations, we have reduced the problem to a fully tractable question with an algorithmic solution. With this, in principle, now the problem can be placed on a computer. In the rest of the paper, we move on to some comments about solving the equation \eqref{main equation} using two different methods. For fermions and bosons carrying only two indices, we are able to find a simple solution to the problem. For higher number of indices, we did not find such a simple approach, but nonetheless we list the classes of states in the antisymmetric case up to level $n=5$ for the three index case. By direct counting, we have verified that the states add up to the expected result for the total number of anti-symmetric states. 

Before discussing our formalism and results further, let us briefly compare our methods with the ones previously studied in the literature (see chapter 7 of \cite{Chen} for example. Also see the first two references in \cite{Coleman}). To describe their approach, let us first label $i_aj_ak_a\equiv I_a$. Then, we see that any fermionic state of the form $\psi ^{i_1j_1k_1}\ldots \psi ^{i_nj_nk_n}\equiv \psi ^{I_1}\ldots \psi ^{I_n}$ is in the completely antisymmetric representation $[1^n]$ of $S_{n_1n_2n_3}$. Then the antisymmetric states of $S_n^{(i)}\times S_n^{(j)}\times S_n^{(k)}$ can be obtained by finding the irreps $a,b,c$ respectively of $S_n^{(i)}, S_n^{(j)}, S_n^{(k)}$ such that the inner product of the irreps $a,b,c$ contains the completely antisymmetric representation. This method involves two steps:
\begin{itemize}
	\item The  Young patterns along with their multiplicities should be identified by using the characters of the irreps under consideration. This step is straightforward although becomes tedious with increasing order of the permutation group.
	\item After that, we need to determine the Clebsch-Gordon coefficients for the inner/Kronecker product of irreps of the permutation group.  Even though there exists some recursion formulas to find these CG coefficients (see first reference in \cite{Coleman} for an example), it is not clear to us whether such a general formula is known. In any event, even if it exists finding these CG coefficients explicitly is in general tedious.  
\end{itemize}
The method we discuss in this paper reformulates the question as an eignevalue problem, and aims to get the complete result in a single shot. Hence, as a byproduct of our formalism to find antisymmetric states, we can read off the CG coefficients of inner product of some of the irreps of $S_n$. But it should be mentioned that for higher levels, our approach also will choke in practice due to the large Young tableaux involved: this seems to be an insurmountable problem because of the factorials involved. 

\section{Two Slots}

Lets start by treating the equation \eqref{main equation} as a set of linear equations and we solve them sequentially starting from $g_1=(12)$ until $g_{n-1}=(n-1,n)$. The number of linear equations are $d_{R_i}d_{R_j}d_{R_c}$ where $d_{R_i}$,$d_{R_j}$ and $d_{R_k}$ are the dimensions of $R_i$, $R_j$ and $R_k$ irreps respectively. We work with Young-Yamanouchi orthonormal basis\footnote{See appendices for some relevant definitions and explanations.} in the rest of the section. 

Before going to the general case, we will attack a simpler problem of finding antisymmetric states of $S_n\otimes S_n$. This corresponds to the case with two quantum numbers. As we show below, we can find a simple solution for this two-index case. But the strategy we employ here takes advantage of specific features limited to this particular case. 

\subsection{Fermions}

The equation that gives us the antisymmetric states of $S_n\times S_n$  is:
\begin{align}
\label{main equation-2 slot}
D(g)_a\otimes D(g)_b\left[\sum_{i,j} \alpha _{ij}^{(a,b)} ~|i\rangle _a\otimes |j\rangle _b\right]&=-\sum_{i,j} \alpha _{ij}^{(a,b)} ~|i\rangle _a\otimes |j\rangle _b
\end{align}
where $a$ and $b$ label the representations of the first and second $S_n$'s respectively. $g_i$ is one of the transpositions (2-cycles) of the form $(i,i+1)$ for $i=1,\ldots (n-1)$.  $D(g)_{a}$ and $D(g)_b$ are the matrix forms of $g$ in the representations $a$ and $b$ of $S_n$ respectively. 

Now, we take an inner product with some specific basis state of the form $|i'\rangle _a\otimes |j'\rangle _b$ to obtain the following:
\begin{align}
\label{matrix element equation}
\sum _{i,j} \alpha _{ij}^{(a,b)}~_a\langle i'|D(g)_a|i\rangle _a~_b\langle j'|D(g)_b|j\rangle _b&=-\alpha _{i'j'}^{(a,b)}
\end{align}
The action of $D(g)$ on the states $|i'\rangle $ and $|j'\rangle $ is as follows:
\begin{align}
D(g)_a|i'\rangle _a&=-p_a^{i'}(g)|i'\rangle _a+\sqrt{1-(p_a^{i'}(g))^2}~|i''\rangle _a \nonumber \\
D(g)_a|i''\rangle _a&=+\sqrt{1-(p_a^{i'}(g))^2}~|i'\rangle _a+p_a^{i'}(g)|i''\rangle _a
\end{align} 
where $|i''\rangle _a$ is another standard Young tableaux that is obtained by exchanging $i$ and $(i+1)$ in $|i'\rangle _a$ when we are working with $g=(i,i+1)$.. Here $p_a^{i'}(g)$ is the axial distance\footnote{It is given by the inverse of number of steps to reach $(i+1)$ starting from $i$ in a standard Young tableaux. So, $p_a^{i'}(g)$ is of the form (positive integer)$^{-1}$. } between $i$ and $(i+1)$ when $g=(i,i+1)$. The basic point about the above equations is that for a given 2-cycle $g$ they mix only two of the states $|i'\rangle _a$ and $|i''\rangle _a$. Similarly, we have:
\begin{align}
D(g)_b|j'\rangle _a&=-p_b^{j'}(g)|j'\rangle _b+\sqrt{1-(p_b^{j'}(g))^2}~|j''\rangle _b \nonumber \\
D(g)_b|j''\rangle _a&=+\sqrt{1-(p_b^{j'}(g))^2}~|j'\rangle _b+p_b^{j'}(g)|j''\rangle _b
\end{align} 
Substituting these expressions in \eqref{matrix element equation} and noting that $|i\rangle$'s form an orthonormal basis, we get:
\begin{align}
\label{eqn-1}
p_a^{i'}(g)p_b^{j'}(g)~\alpha ^{(a,b)}_{i'j'}-\sqrt{1-(p_a^{i'}(g))^2}p_b^{j'}(g)~\alpha ^{(a,b)}_{i''j'}-p_a^{i'}(g) \sqrt{1-(p_b^{j'}(g))^2}~\alpha ^{(a,b)}_{i'j''}&\nonumber \\ 
+\sqrt{1-(p_a^{i'}(g))^2}\sqrt{1-(p_b^{j'}(g))^2}~\alpha ^{(a,b)}_{i''j''}&=-\alpha ^{(a,b)}_{i'j'}
\end{align}

Taking an inner product with $|i''\rangle _a\otimes |j'\rangle _b$, we get:
\begin{align}
\label{eqn-2}
-\sqrt{1-(p_a^{i'}(g))^2}p_b^{j'}(g)~\alpha ^{(a,b)}_{i'j'}-p_a^{i'}(g)p_b^{j'}(g)~\alpha ^{(a,b)}_{i''j'}&\nonumber \\ +\sqrt{1-(p_a^{i'}(g))^2} \sqrt{1-(p_b^{j'}(g))^2}~\alpha ^{(a,b)}_{i'j''}
+p_a^{i'}(g)\sqrt{1-(p_b^{j'}(g))^2}~\alpha ^{(a,b)}_{i''j''}&=-\alpha ^{(a,b)}_{i''j'}
\end{align}

Taking an inner product with $|i'\rangle _a\otimes |j''\rangle _b$, we get:
\begin{align}
\label{eqn-3}
-p_a^{i'}(g)\sqrt{1-(p_b^{j'}(g))^2}~\alpha ^{(a,b)}_{i'j'}+\sqrt{1-(p_a^{i'}(g))^2} \sqrt{1-(p_b^{j'}(g))^2}~\alpha ^{(a,b)}_{i''j'} &\nonumber \\ -p_a^{i'}(g)p_b^{j'}(g)~\alpha ^{(a,b)}_{i'j''} 
+\sqrt{1-(p_a^{i'}(g))^2}p_b^{j'}(g)~\alpha ^{(a,b)}_{i''j''}&=-\alpha ^{(a,b)}_{i'j''}
\end{align}

Taking an inner product with $|i''\rangle _a\otimes |j''\rangle _b$, we get:
\begin{align}
\label{eqn-4}
\sqrt{1-(p_a^{i'}(g))^2}\sqrt{1-(p_b^{j'}(g))^2}~\alpha ^{(a,b)}_{i'j'}+p_a^{i'}(g) \sqrt{1-(p_b^{j'}(g))^2}~\alpha ^{(a,b)}_{i''j'} &\nonumber \\  +\sqrt{1-(p_a^{i'}(g))^2}p_b^{j'}(g)~\alpha ^{(a,b)}_{i'j''}
+p_a^{i'}(g)p_b^{j'}(g)~\alpha ^{(a,b)}_{i''j''}&=-\alpha ^{(a,b)}_{i''j''}
\end{align}
Our goal is to find constraints between $p_a^{i'}(g)$ and $p_b^{j'}(g)$ so that the above four equations have a non-trivial solution for $\alpha $'s. Before proceeding further, we recall that the $\alpha $'s are independent of the 2-cycle $g$.

%\subsection{Sequential filling of Young Tableaux in 2-slot case}

By solving \eqref{matrix element equation}, we obtain $\alpha $'s in terms of $p_a$ and $p_b$. But, our aim is to constrain $p_a$ and $p_b$ themselves using the equations  \eqref{matrix element equation}. We can possibly obtain such constraints by demanding the existence of non-trivial solutions to the equations \eqref{eqn-1}-\eqref{eqn-4}. If we write the equation \eqref{eqn-1}-\eqref{eqn-4} as $Ax=0$ schematically, then this is same as demanding that the determinant of $A$ is zero. But, it can be checked that the determinant of $A$ is trivially zero if all the $\alpha$'s are nonzero. So, it seems that we cannot constrain $p_a$ and $p_b$.

But, we now argue that we can indeed constrain $p_a^{(g)}$ and $p_b^{(g)}$. For any given $g_i$, we claim that only two of the four $\alpha $'s that occur in equations \eqref{eqn-1}-\eqref{eqn-4} are non-zero because of the constraints imposed by $g_1\ldots g_{i-1}$. Before giving evidence to support the claim, we discuss its implications. Once we accept the claim, by demanding that the $\alpha $'s have a non-trivial solution, we get\footnote{The $\pm$ sign depends on our choices of Young tableaux.} $p_a^{(g)}=\pm p_b^{(g)}$. This condition translates to the statement that the distance between $i$ and $(i+1)$ in the Young tableaux in first slot is equal to positive/negative\footnote{The distance is equal to the number of steps to reach from $i$ to $(i+1)$. When the steps are counted left or down, we take them to be positive distance. Right or upward steps contribute to negative distance. Further, note that the positions of $i$ and $i+1$ are exchanged in $|i'\rangle $ to obtain $|i''\rangle $.} of the distance between $i$ and $(i+1)$ in the second Young tableaux. Equivalently, we start by filling 1 and 2 and then pick a spot for 3 in the first tableaux and then the position of 3 in the second tableaux is fixed by the above condition of distances. We continue this process to obtain the entire tableaux in the second slot corresponding to a tableaux in the first slot and this solves the problem. 

But this argument depends on the uniqueness of the second tableaux for a given tableaux in the first slot. We will not prove this statement, but we have checked that it is true for the first six levels, and we present some of the details below. We believe this is true generally.

Suppose we have filled the Young tableaux in both the slots from 1 to $i$ such that it is a part of antisymmetric state. We now want to fill the $(i+1)$ in both the tableaux so that it forms a part of antisymmetric state. If we fix the position of $(i+1)$ in the first tableaux, then we have (at most) four states that are obtained by permuting the $i$ and $(i+1)$ indices. We have represented these four states in the previous section as:
\begin{align}
|i'\rangle \otimes |j'\rangle ; \hspace{5 mm} |i''\rangle \otimes |j'\rangle ; \hspace{5 mm} |i'\rangle \otimes |j''\rangle ; \hspace{5 mm} |i''\rangle \otimes |j''\rangle 
\end{align}
Note that while permuting $i$ and $(i+1)$, we do change the distance between $(i-1)$ and $i$. As a result, for some fixed positions of $(i-1)$ and $i$ in the first Young tableaux, we have two different sets of positions of $(i-1)$ and $i$ in the second Young tableaux. But below, we will see that only one of them in fact appears at low levels.\\
%we give examples proving that for a Young tableaux in the first slot, there is a unique Young tableaux in the second slot.

Let us start with $g_1=(12)$. In this case, there are only two antisymmetric states at level 2 and they are given by:
\begin{figure}[htbp!]
	\centering
	\includegraphics[trim={4cm 25.4cm 3.7cm 2.5cm},clip,scale=1]{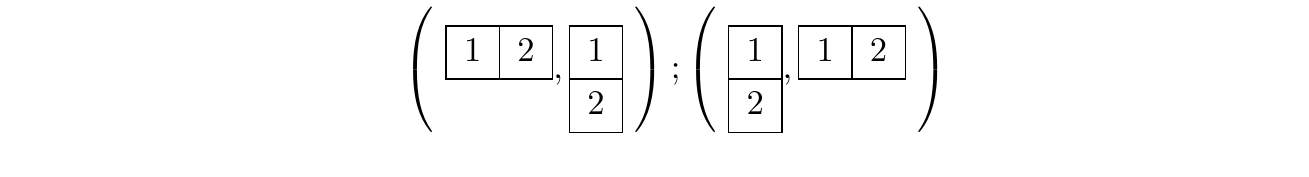}
\end{figure}\\
%\begin{align}
%\left(~\begin{ytableau}1&2 \end{ytableau},\begin{ytableau}1\\2 %\end{ytableau}~\right); \left(~\begin{ytableau}1\\2 %\end{ytableau},\begin{ytableau}1&2 \end{ytableau}~\right)
%\end{align}
Now we move on to the Young tableaux involving three boxes. It is easy to check that $p_a=\pm 1; p_b=\pm 1$ satisfies the equations \eqref{eqn-1}-\eqref{eqn-4}.That is, following are antisymmetric states:
\begin{figure}[htbp!]
	\centering
	\includegraphics[trim={4cm 24.2cm 3.7cm 2.5cm},clip,scale=1]{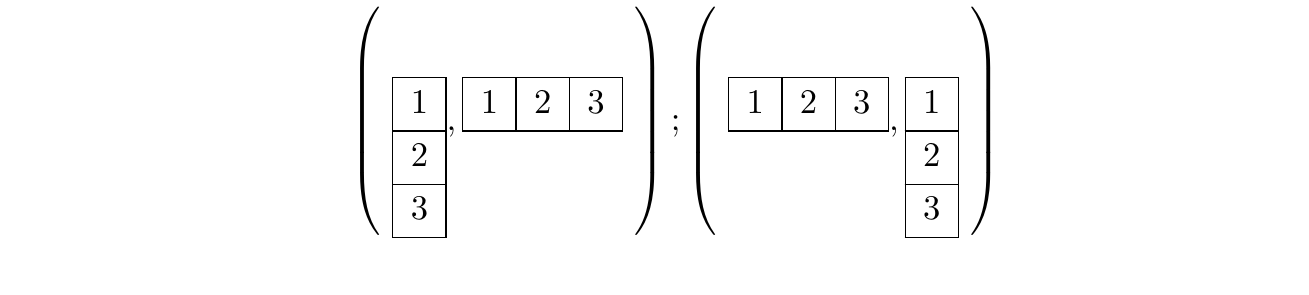}
\end{figure}\\
%\begin{align}
 %\left(~\begin{ytableau}1\\2\\3 \end{ytableau},\begin{ytableau}1&2&3 %\end{ytableau}~\right); \left(~\begin{ytableau}1&2&3 %\end{ytableau},\begin{ytableau}1\\2\\3 %\end{ytableau}~\right)\nonumber \\
%\end{align} 
Consider the following states corresponding to the mixed symmetry Young tableau:
\begin{figure}[htbp!]
	\centering
	\includegraphics[trim={4cm 25.5cm 3cm 2.5cm},clip,scale=1]{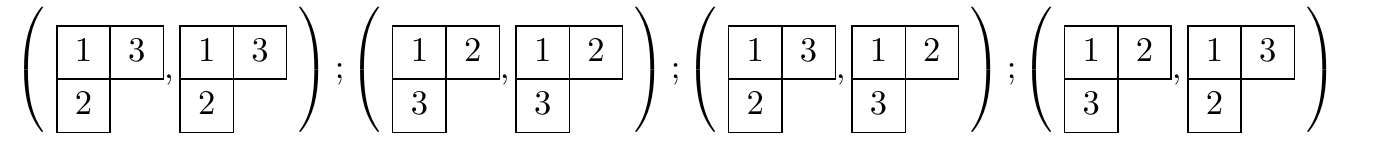}
\end{figure} \\
%\begin{align}
 %\left(~\begin{ytableau}1&3\\2 \end{ytableau},\begin{ytableau}1&3\\2 \end{ytableau}~\right); \left(~\begin{ytableau}1&2\\3 \end{ytableau},\begin{ytableau}1&2\\3 \end{ytableau}~\right);  \left(~\begin{ytableau}1&3\\2 \end{ytableau},\begin{ytableau}1&2\\3 \end{ytableau}~\right); \left(~\begin{ytableau}1&2\\3 \end{ytableau},\begin{ytableau}1&3\\2 \end{ytableau}~\right)
%\end{align}
\newline
\noindent We can choose $i'$ and $j'$ to be mixed symmetric tableaux with 1,2 in the same row. Then from $g_1=(12)$ we see that $\alpha _{i''j''}=0=\alpha _{i'j'}$. So, only two $\alpha $'s survive as expected. The antisymmetric state under $(12)$ and $(23)$ is:
\begin{figure}[htbp!]
	\centering
	\includegraphics[trim={4cm 25.5cm 3cm 2.5cm},clip,scale=1]{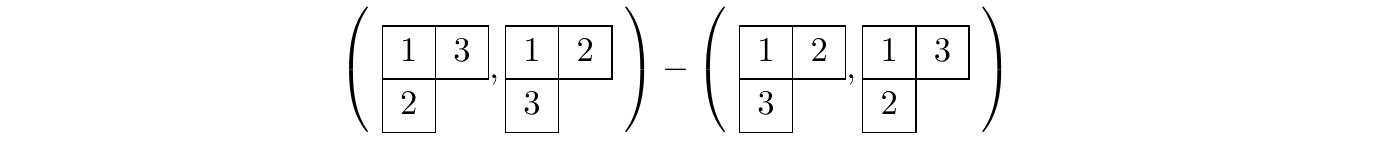}
\end{figure}\\
%\begin{align}
%\label{3-level antisymmetric}
 %\left(~\begin{ytableau}1&3\\2 \end{ytableau},\begin{ytableau}1&2\\3 \end{ytableau}~\right)-\left(~\begin{ytableau}1&2\\3 \end{ytableau},\begin{ytableau}1&3\\2 \end{ytableau}~\right)
%\end{align} 
We can see that the Young tableaux in the second slot is unique with respect to the Young tableaux in the first slot.

As a level 4 example, consider the following set of states:
\begin{figure}[htbp!]
	\centering
	\includegraphics[trim={4cm 22cm 3cm 2.5cm},clip,scale=1]{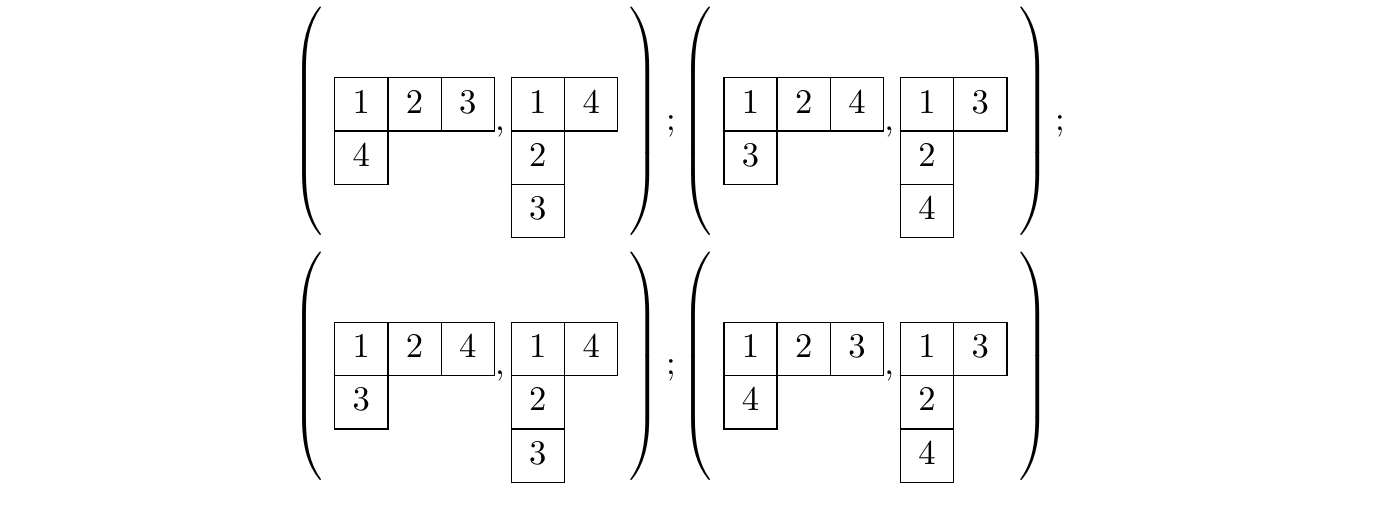}
\end{figure}\\
%\begin{align}
%&\left(~\begin{ytableau}1&2&3\\4 \end{ytableau},\begin{ytableau}1&4\\2\\3 \end{ytableau}~\right);\left(~\begin{ytableau}1&2&4\\3 \end{ytableau},\begin{ytableau}1&3\\2\\4 \end{ytableau}~\right);\non\\
%&\left(~\begin{ytableau}1&2&4\\3 \end{ytableau},\begin{ytableau}1&4\\2\\3 \end{ytableau}~\right);\left(~\begin{ytableau}1&2&3\\4 \end{ytableau},\begin{ytableau}1&3\\2\\4 \end{ytableau}~\right)
%\end{align}  
We choose $i'$ and $j'$ to be the Young tableaux such that 1,2,3 are in the same row in $i'$ and in the same column in $j'$.  From $g_2=(23)$, we can see that only $\alpha _{i'j''}$ and $\alpha _{i''j'}$ are non-zero. The antisymmetric state under $(34)$ is:
\begin{figure}[h!]
	\centering
	\includegraphics[trim={4cm 24.5cm 3cm 2.5cm},clip,scale=1]{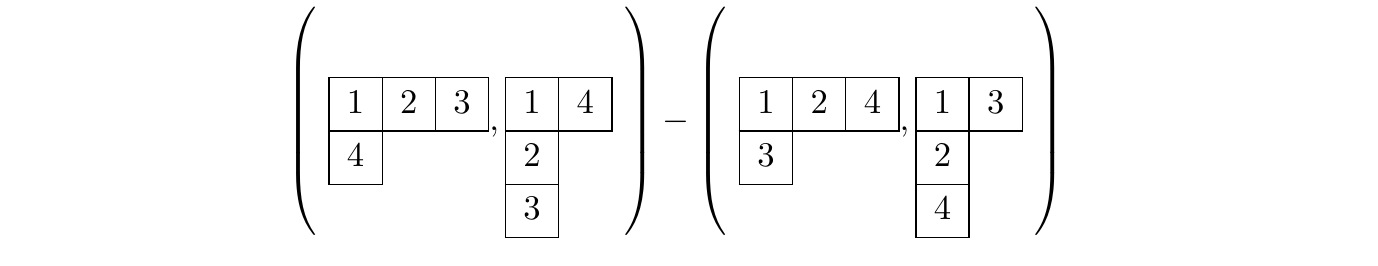}
\end{figure}
%\begin{align}
%\left(~\begin{ytableau}1&2&3\\4 \end{ytableau},\begin{ytableau}1&4\\2\\3 \end{ytableau}~\right)-\left(~\begin{ytableau}1&2&4\\3 \end{ytableau},\begin{ytableau}1&3\\2\\4 \end{ytableau}~\right)
%\end{align}
 
\noindent For it to be a antisymmetric state, we need to add one more term following the level-3 antisymmetric states and it is given by:
 \begin{figure}[h!]
 	\centering
 	\includegraphics[trim={4cm 24.5cm 3cm 2.5cm},clip,scale=1]{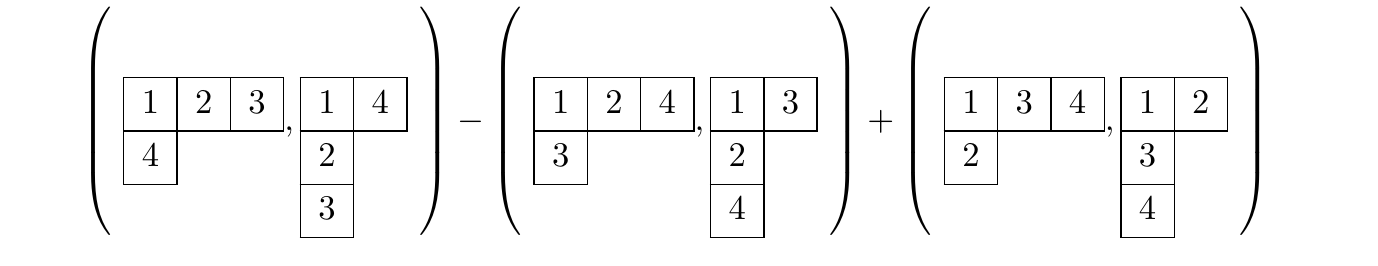}
 \end{figure}\\
 %\begin{align}
 %\left(~\begin{ytableau}1&2&3\\4 \end{ytableau},\begin{ytableau}1&4\\2\\3 \end{ytableau}~\right)-\left(~\begin{ytableau}1&2&4\\3 \end{ytableau},\begin{ytableau}1&3\\2\\4 \end{ytableau}~\right)+\left(~\begin{ytableau}1&3&4\\2 \end{ytableau},\begin{ytableau}1&2\\3\\4 \end{ytableau}~\right)
 %\end{align}
Here again, the Young tableaux in the second slot is unique with respect to the Young tableaux in the first slot.  We have checked this uniqueness explicitly for  Young tableaux with up to six boxes (ie., level 6) and we expect it to be true in general. Also, the solution obtained here passes the counting check we have described in the final section.

\subsection{Bosons}

As a simple corollary of our approach, we can construct baryonic wave functions. As explained in the introduction, this requires us to consider the bosonic case. In this subsection, we make a digression to do so. The results are quite parallel to the two-slot fermionic case. %In constructing these wave functions, we deal with three quantum numbers corresponding to spin, flavor and color. Since we know that the complete baryonic wave function should be antisymmetric and also find only color singlets in nature, this implies that the wave function should be symmetric under the spin and flavor quantum numbers. In this section, 
We present a solution to a generalized version of this problem i.e., we find the fully symmetric representations of the group $G_i\times G_j$ where we take $G_{i,j}$ to be $U(n_{i,j})$ for concreteness. More operationally, we solve our main equation \eqref{main equation} for two index case with a $+$ sign on the RHS. In the rest of the section, we work with the Young-Yamanouchi representation.

The equation that gives us the symmetric states of $S_n\times S_n$ is given by:
\begin{align}
\label{2-slot symmetric}
D(g)_a\otimes D(g)_b\left[\sum_{i,j} \alpha _{ij}^{(a,b)} ~|i\rangle _a\otimes |j\rangle _b\right]&=+\sum_{i,j} \alpha _{ij}^{(a,b)} ~|i\rangle _a\otimes |j\rangle _b
\end{align}
where $a$ and $b$ are certain representations of the first and second $S_n$'s respectively. $g$ is one of the transpositions (2-cycles) of the form $(i,i+1)$ for $i=1,\ldots (n-1)$.  $D(g)_{a}$ and $D(g)_b$ are the matrix forms of $g$ in the representations $a$ and $b$ of $S_n$ respectively. 

We now take an inner product on both sides of the equation \eqref{2-slot symmetric} with a specific basis state $|i'\rangle _a \otimes |j'\rangle _b$  to obtain the following:
\begin{align}
\label{eqn-1:symmetric}
p_a^{i'}(g)p_b^{j'}(g)~\alpha ^{(a,b)}_{i'j'}-\sqrt{1-(p_a^{i'}(g))^2}p_b^{j'}(g)~\alpha ^{(a,b)}_{i''j'}-p_a^{i'}(g) \sqrt{1-(p_b^{j'}(g))^2}~\alpha ^{(a,b)}_{i'j''}&\nonumber \\ 
+\sqrt{1-(p_a^{i'}(g))^2}\sqrt{1-(p_b^{j'}(g))^2}~\alpha ^{(a,b)}_{i''j''}&=+\alpha ^{(a,b)}_{i'j'}
\end{align}
where $|i''\rangle _a$ is another standard Young tableaux that is obtained by exchanging $i$ and $(i+1)$ in $|i'\rangle _a$ when we are working with $g=(i,i+1)$. $p_a^{i'}(g)$ is the axial distance between $i$ and $(i+1)$.

Taking inner products with $|i''\rangle _a \otimes |j'\rangle _b$, $|i'\rangle _a \otimes |j''\rangle _b$ and $|i''\rangle _a \otimes |j''\rangle _b$ gives the following equations:
\begin{align}
\label{eqn-2:symmetric}
-\sqrt{1-(p_a^{i'}(g))^2}p_b^{j'}(g)~\alpha ^{(a,b)}_{i'j'}-p_a^{i'}(g)p_b^{j'}(g)~\alpha ^{(a,b)}_{i''j'} &\nonumber \\ +\sqrt{1-(p_a^{i'}(g))^2} \sqrt{1-(p_b^{j'}(g))^2}~\alpha ^{(a,b)}_{i'j''}
+p_a^{i'}(g)\sqrt{1-(p_b^{j'}(g))^2}~\alpha ^{(a,b)}_{i''j''}&=+\alpha ^{(a,b)}_{i''j'}\\
\label{eqn-3:symmetric}
-p_a^{i'}(g)\sqrt{1-(p_b^{j'}(g))^2}~\alpha ^{(a,b)}_{i'j'}+\sqrt{1-(p_a^{i'}(g))^2} \sqrt{1-(p_b^{j'}(g))^2}~\alpha ^{(a,b)}_{i''j'} &\nonumber \\  -p_a^{i'}(g)p_b^{j'}(g)~\alpha ^{(a,b)}_{i'j''}
+\sqrt{1-(p_a^{i'}(g))^2}p_b^{j'}(g)~\alpha ^{(a,b)}_{i''j''}&=+\alpha ^{(a,b)}_{i'j''}\\
\label{eqn-4:symmetric}
\sqrt{1-(p_a^{i'}(g))^2}\sqrt{1-(p_b^{j'}(g))^2}~\alpha ^{(a,b)}_{i'j'}+p_a^{i'}(g) \sqrt{1-(p_b^{j'}(g))^2}~\alpha ^{(a,b)}_{i''j'} &\nonumber \\  +\sqrt{1-(p_a^{i'}(g))^2}p_b^{j'}(g)~\alpha ^{(a,b)}_{i'j''}
+p_a^{i'}(g)p_b^{j'}(g)~\alpha ^{(a,b)}_{i''j''}&=+\alpha ^{(a,b)}_{i''j''}
\end{align}
As in the case of anti-symmetrizing states, we claim that the Young tableaux in the second slot is unique for a given Young tableaux in the first slot. This claim about uniqueness implies that we need\footnote{The $\pm$ sign depends on our choices of Young tableaux.} $p_a^{i'}(g)=\pm p_b^{j'}(g)$ so that the equations \eqref{eqn-1:symmetric}-\eqref{eqn-4:symmetric} have a non-trivial solution.

Now, we give examples to support the above claim. At level 2, we have only two symmetric representations and are given by:
\begin{figure}[h!]
	\centering
	\includegraphics[trim={4cm 25.5cm 3cm 2.5cm},clip,scale=1]{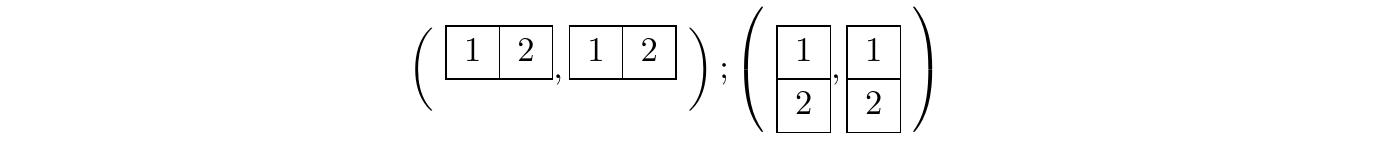}
\end{figure}\\
%\begin{align}
%\left(~\begin{ytableau}1&2 \end{ytableau},\begin{ytableau}1&2 \end{ytableau}~\right); \left(~\begin{ytableau}1\\2 \end{ytableau},\begin{ytableau}1\\2 \end{ytableau}~\right)
%\end{align}
Moving on to level 3, we have the following symmetric representations:
\begin{figure}[h!]
	\centering
	\includegraphics[trim={4cm 24.5cm 3cm 2.5cm},clip,scale=1]{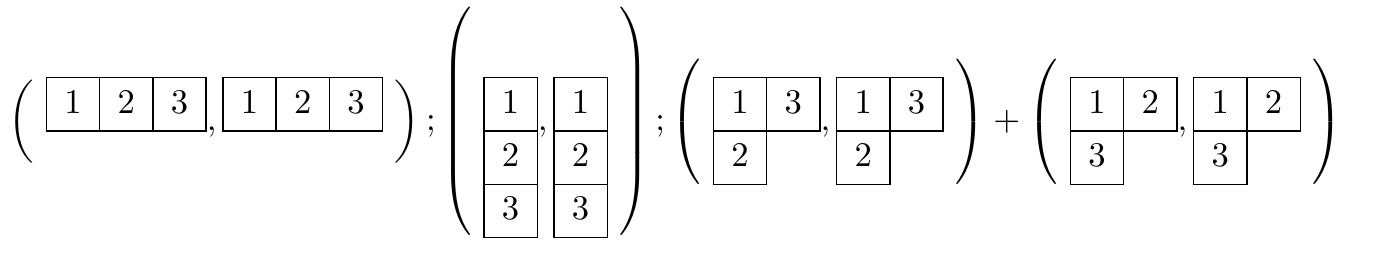}
\end{figure}\\
%\begin{align}
%\left(~\begin{ytableau}1&2&3 \end{ytableau},\begin{ytableau}1&2&3 \end{ytableau}~\right); \left(~\begin{ytableau}1\\2\\3 \end{ytableau},\begin{ytableau}1\\2\\3 \end{ytableau}~\right);\left(~\begin{ytableau}1&3\\2 \end{ytableau},\begin{ytableau}1&3\\2 \end{ytableau}~\right)+\left(~\begin{ytableau}1&2\\3 \end{ytableau},\begin{ytableau}1&2\\3 \end{ytableau}~\right)
%\end{align}	
The discussion is parallel to the anti-symmetric case, so we will not belabor it. We have checked this uniqueness up to level-6 and we expect it to work at an arbitrary level. Also, the symmetric representations we obtain here passes the counting check that we describe in the final section. 

The results we find here, when interpreted as flavor and spin quantum numbers provides the solution to the baryon wave function problem. The solution to this problem for the case of three ($u, d, s$) flavors can be found in \cite{Griffiths} for example.

\section{Three Slots}

In the case with three kinds of quantum numbers, we will stick to the fermionic case. The bosonic case is analogous, but since there is no immediate physical application we have in mind (unlike the baryon wave functions in the two slot case) we will not spell it out explicitly.

The equation that we intend to solve to obtain the antisymmetric states in 3-index case is given by:
\begin{align}
%\label{main equation-3 slot}
D(g)_a\otimes D(g)_b\otimes D(g)_c\left[\sum_{i,j,k} \alpha _{ijk}^{(a,b,c)} ~|i\rangle _a\otimes |j\rangle _b\otimes |k\rangle _c\right]&=-\sum_{i,j,k} \alpha _{ijk}^{(a,b,c)} ~|i\rangle _a\otimes |j\rangle _b\otimes |k\rangle _c
\end{align}

Taking an inner product with a certain basis state $|i'\rangle _a\otimes |j'\rangle _b\otimes |k'\rangle _c$, we get:
\begin{align}
\label{matrix 3-slot}
\sum_{i,j,k} \alpha _{ijk}^{(a,b,c)} \langle i'|D(g)_a|i\rangle _a ~\langle j'|D(g)_b|j\rangle _b ~\langle k'|D(g)_c|k\rangle _c &=-\alpha _{i'j'k'}^{(a,b,c)}
\end{align}
As in the last section, we take:
\begin{align}
D(g)_a|i'\rangle _a&=-p_a^{i'}(g)|i'\rangle _a+\sqrt{1-(p_a^{i'}(g))^2}~|i''\rangle _a \nonumber \\
D(g)_a|i''\rangle _a&=+\sqrt{1-(p_a^{i'}(g))^2}~|i'\rangle _a+p_a^{i'}(g)|i''\rangle _a	\\
D(g)_b|j'\rangle _b&=-p_b^{j'}(g)|j'\rangle _b+\sqrt{1-(p_b^{j'}(g))^2}~|j''\rangle _b \nonumber \\
D(g)_b|j''\rangle _b&=+\sqrt{1-(p_b^{j'}(g))^2}~|j'\rangle _b+p_b^{j'}(g)|j''\rangle _b \\
D(g)_c|k'\rangle _c&=-p_c^{k'}(g)|k'\rangle _c+\sqrt{1-(p_c^{k'}(g))^2}~|k''\rangle _c \nonumber \\
D(g)_c|k''\rangle _c&=+\sqrt{1-(p_c^{k'}(g))^2}~|k'\rangle _c+p_c^{k'}(g)|k''\rangle _c
\end{align}
where $|i''\rangle _a$, $|j''\rangle _b$ and $|k''\rangle _c$ are the basis states of the $a$, $b$ and $c$ irreps respectively. Putting these expressions back into \eqref{matrix 3-slot}, we get:
\begin{align}
-\alpha _{i'j'k'}=&-p_a  \left[\alpha _{i'j'k'}~p_bp_c-\alpha _{i'j''k'}~\sqrt{1-p^2_b}~p_c \right.\non\\
&\qquad\qquad \left.  -\alpha _{i'j'k''}~p_b\sqrt{1-p^2_c}+\alpha _{i'j''k''}~\sqrt{1-p^2_b}\sqrt{1-p_c^2}\right] \nonumber \\
&+\sqrt{1-p_a^2}\left[\alpha _{i''j'k'}~p_bp_c-\alpha _{i''j''k'}~\sqrt{1-p^2_b}~p_c \right.\non\\
&\qquad\qquad\qquad \left.-\alpha _{i''j'k''}~p_b\sqrt{1-p^2_c}+\alpha _{i''j''k''}~\sqrt{1-p^2_b}\sqrt{1-p_c^2}\right]
\end{align}
where we have dropped various subscripts and superscripts to avoid clutter of notation. In a similar way, we get seven more equations by taking inner product with various states and those equations can be listed as follows:
\begin{align}
\label{3-slot equations}
-\alpha _{i''j'k'}=& \sqrt{1-p_a^2}\left[\alpha _{i'j'k'}~p_bp_c-\alpha _{i'j''k'}~\sqrt{1-p^2_b}~p_c\right.\non\\
&\qquad\qquad \left. -\alpha _{i'j'k''}~p_b\sqrt{1-p^2_c}+\alpha _{i'j''k''}~\sqrt{(1-p^2_b)(1-p_c^2)}\right]\nonumber \\
&+p_a\left[\alpha _{i''j'k'}~p_bp_c-\alpha _{i''j''k'}~\sqrt{1-p^2_b}~p_c \right.\non\\
&\qquad\qquad \left.-\alpha _{i''j'k''}~p_b\sqrt{1-p^2_c}+\alpha _{i''j''k''}~\sqrt{1-p^2_b}\sqrt{1-p_c^2}\right]\\
-\alpha _{i'j''k'}=&-p_a\left[-\alpha _{i'j'k'}~\sqrt{1-p^2_b}~p_c-\alpha _{i'j''k'}~p_b~p_c \right.\non\\
&\qquad\qquad \left. +\alpha _{i'j'k''}~\sqrt{1-p^2_b}\sqrt{1-p^2_c}+\alpha _{i'j''k''}~p_b\sqrt{1-p_c^2}\right]\nonumber \\
&+\sqrt{1-p_a^2}\left[-\alpha _{i''j'k'}~\sqrt{1-p^2_b}~p_c-\alpha _{i''j''k'}~p_b~p_c \right.\non\\
&\qquad\qquad\qquad \left. +\alpha _{i''j'k''}~\sqrt{1-p^2_b}\sqrt{1-p^2_c}+\alpha _{i''j''k''}~p_b\sqrt{1-p_c^2}\right]\\
-\alpha _{i''j''k'}=&\sqrt{1-p_a^2}\left[-\alpha _{i'j'k'}~\sqrt{1-p^2_b}~p_c-\alpha _{i'j''k'}~p_b~p_c \right.\non\\
&\qquad\qquad\qquad \left. +\alpha _{i'j'k''}~\sqrt{1-p^2_b}\sqrt{1-p^2_c}+\alpha _{i'j''k''}~p_b\sqrt{1-p_c^2}\right]\nonumber \\
&+p_a\left[-\alpha _{i''j'k'}~\sqrt{1-p^2_b}~p_c-\alpha _{i''j''k'}~p_b~p_c\right.\non\\
&\qquad\qquad\qquad \left. + \alpha _{i''j'k''}~\sqrt{1-p^2_b}\sqrt{1-p^2_c}+\alpha _{i''j''k''}~p_b\sqrt{1-p_c^2}\right] \\
-\alpha _{i'j'k''}=& -p_a\left[-\alpha _{i'j'k'}~p_b\sqrt{1-p_c^2}+\alpha _{i'j''k'}~\sqrt{1-p^2_b}~\sqrt{1-p_c^2} \right.\non\\
&\qquad\qquad\qquad\qquad \left. -\alpha _{i'j'k''}~p_b~p_c +\alpha _{i'j''k''}~\sqrt{1-p^2_b}~p_c\right]\nonumber \\
&+\sqrt{1-p_a^2}\left[-\alpha _{i''j'k'}~p_b\sqrt{1-p_c^2}+\alpha _{i''j''k'}~\sqrt{1-p^2_b}~\sqrt{1-p_c^2} \right.\non\\
&\qquad\qquad\qquad\qquad\qquad \left.  -\alpha _{i''j'k''}~p_b~p_c+\alpha _{i''j''k''}~\sqrt{1-p^2_b}~p_c\right]\\
-\alpha _{i''j'k''}=&\sqrt{1-p_a^2}\left[-\alpha _{i'j'k'}~p_b\sqrt{1-p_c^2}+\alpha _{i'j''k'}~\sqrt{1-p^2_b}~\sqrt{1-p_c^2} \right.\non\\
&\qquad\qquad\qquad\qquad\qquad \left. -\alpha _{i'j'k''}~p_b~p_c+\alpha _{i'j''k''}~\sqrt{1-p^2_b}~p_c\right]\nonumber \\
&+p_a\left[-\alpha _{i''j'k'}~p_b\sqrt{1-p_c^2}+\alpha _{i''j''k'}~\sqrt{1-p^2_b}~\sqrt{1-p_c^2}\right.\non\\
&\qquad\qquad\qquad\qquad \left.  -\alpha _{i''j'k''}~p_b~p_c+\alpha _{i''j''k''}~\sqrt{1-p^2_b}~p_c\right]\end{align}\begin{align}
-\alpha _{i'j''k''}=&-p_a\left[\alpha _{i'j'k'}~\sqrt{1-p_b^2}\sqrt{1-p_c^2}+\alpha _{i'j''k'}~p_b\sqrt{1-p_c^2}\right.\non\\
&\qquad\qquad\qquad\qquad \left.  +\alpha _{i'j'k''}~\sqrt{1-p_b^2}~p_c+\alpha _{i'j''k''}~p_bp_c\right]\nonumber \\
&+\sqrt{1-p_a^2}\left[\alpha _{i''j'k'}~\sqrt{1-p_b^2}\sqrt{1-p_c^2}+\alpha _{i''j''k'}~p_b\sqrt{1-p_c^2}\right.\non\\
&\qquad\qquad\qquad\qquad\qquad \left.  +\alpha _{i''j'k''}~\sqrt{1-p_b^2}~p_c+\alpha _{i''j''k''}~p_bp_c\right]\\
-\alpha _{i''j''k''}=&\sqrt{1-p_a^2}\left[\alpha _{i'j'k'}~\sqrt{1-p_b^2}\sqrt{1-p_c^2}+\alpha _{i'j''k'}~p_b\sqrt{1-p_c^2} \right.\non\\
&\qquad\qquad\qquad\qquad\qquad \left. +\alpha _{i'j'k''}~\sqrt{1-p_b^2}~p_c+\alpha _{i'j''k''}~p_bp_c\right]\nonumber \\
&+p_a\left[\alpha _{i''j'k'}~\sqrt{1-p_b^2}\sqrt{1-p_c^2}+\alpha _{i''j''k'}~p_b\sqrt{1-p_c^2}\right.\non\\
&\qquad\qquad\qquad\qquad\qquad \left. +\alpha _{i''j'k''}~\sqrt{1-p_b^2}~p_c+\alpha _{i''j''k''}~p_bp_c\right]
\end{align}

Just as in the two slot case, one can check that if all the $\alpha$'s are assumed to be non-vanishing, demanding nontrivial solutions to these equations via a determinant condition does not constrain $p_a$, $p_b$ and $p_c$. But unlike in the two slot case, we have not found a simple approach to setting certain $\alpha$'s to zero that leads to a useful way to enumerate the solutions. At a practical level, this is because for given tableaux in the first two slots, the tableau in the third slot need  not be unique. Of course, one can solve these equations by explicit calculation, and in an appendix, we give (examples of) antisymmetric states at levels  2, 3 and 4.

But we can proceed further by approaching the problem from a different angle, and that is what we turn to next. This alternate approach gives a fairly simple way to find the form of the Young {\em patterns} that show up in the anti-symmetric states\footnote{If we wish, we can further go ahead and solve the problem more completely, and find the answer that includes the precise combinations of Young {\em tableaux} that show up in the Young patterns. But to present the answer in a reasonably compact form, we limit ourselves to just listing the patterns in the Appendix.}. Note that all the specific statements we are making in this section and the last are specific simplifications, we do not claim absolute generality with these methods (beyond the fact that the original equations themselves yield an eigenvalue problem which is obviously tractable with infinite computing power).

\subsection{Auxiliary Eigenvalue Problems}

In this section, we treat the equations \eqref{main equation} as a set of $(n-1)$ eigenvalue equations. Our goal is to find eigenvector(s) (corresponding to eigenvalue of -1) that is common to all the $(n-1)$ matrices of the form $D^a(g_i)\otimes D^b(g_i)\otimes D^c(g_i)$. Here $g_i$ is a group element of $S_n$ and denotes a 2-cycle of the form $(i,i+1)$ where $i$ runs from 1 to $(n-1)$. The superscripts $a,b,c$ denote the particular irreducible representations of $S_n$ we are dealing with.

As in the previous section, we work with Young-Yamanouchi orthonormal representation and each of the standard Young tableaux are given by column matrices of the form:
\begin{align*}
\begin{bmatrix}
~1~ \\ 0 \\ \vdots \\ 0
\end{bmatrix};
\begin{bmatrix}
~0~ \\ 1 \\ \vdots \\ 0
\end{bmatrix};\hdots
\begin{bmatrix}
~0~ \\ 0 \\ \vdots \\ 1
\end{bmatrix}
\end{align*}
In this representation, the matrices $D(g_i)$ corresponding to the transpositions $g_i$ are given by a simple form as explained in an appendix. As each $D(g_i)$ squares to 1, the eigenvalues are $\pm 1$. Thanks to the structure of these matrices, the eigenvectors of each $D(g_i)$ are also easy to write down explicitly. Note that the general structure of these matrices is  given by:
\begin{align}
\begin{bmatrix}
1\\
&  \ddots \\
& & -1 \\
& &  & \ddots \\
& & & &  -\cos \theta _1 & \sin \theta _1 \\
& & & &   \sin \theta _1 & \cos \theta _1 \\
& & & & & & \ddots \\
& & & & & & & -\cos \theta _2 & \sin \theta _2 \\
& & & & & & & \sin \theta _2 & \cos \theta _2 \\
& & & & & & & & & \ddots 
\end{bmatrix}
\end{align}
where $\cos \theta_i \equiv \rho _i$ is the inverse distance that appeared in the previous sections. See appendix \ref{yamanouchi basis} for more details on how to construct the Young-Yamanouchi representation. The eigenvectors are straightforward to obtain. For instance, the eigenvectors corresponding to the eigenvalue $-1$ can be written as:
\begin{align}
\begin{bmatrix}
0 \\ \vdots \\ 1 \\ \vdots \\ 1 \\ -\tan \left(\frac{\theta _1}{2}\right) \\ \vdots  \\ \vdots
\end{bmatrix};
\begin{bmatrix}
0 \\ \vdots \\ 1 \\ \vdots \\  \vdots \\ 1 \\ -\tan \left(\frac{\theta _2}{2}\right) \\ \vdots
\end{bmatrix}; \ldots
\end{align} 

Now that we know how to write down the eigenvectors of each $D(g_i)$ in an arbitrary irrep, we use them to construct the eigenvectors of $D^a(g_i)\otimes D^b(g_i)\otimes D^c(g_i)$. If we denote $a_i$, $b_i$ and $c_i$ as eigenvectors of $D^a(g_i)$, $D^b(g_i)$ and $D^c(g_i)$ with eigenvalues $a$, $b$ and $c$ respectively, then it is easy to show that $a_i \otimes b_i \otimes c_i$ is an  eigenvector of $D^a(g_i)\otimes D^b(g_i)\otimes D^c(g_i)$ with eigenvalue ``$abc$''. Note that any eigenvector of   $D^a(g_i)\otimes D^b(g_i)\otimes D^c(g_i)$ can be written as a tensor product of eigenvectors of the individual components. The corresponding eigenvalue would be the product of the corresponding individual eigenvalues.

Our goal is to find eigenvectors of $D^a(g_i)\otimes D^b(g_i)\otimes D^c(g_i)$ that have an eigenvalue of $-1$ and are common to all the $g_i$'s. In general, for each of $D^a(g_i)\otimes D^b(g_i)\otimes D^c(g_i)$, the eigenvalue $-1$ is degenerate and thus the common eigenvector(s) can be a linear combination of eigenvectors corresponding to each $g_i$. That is, if $\alpha ^{(i)}_{p_i}$ denote\footnote{The subscript $p_i$ is supposed to index the degeneracy in the eigenvectors of $D^a(g_i)\otimes D^b(g_i)\otimes D^c(g_i)$. The superscript $(i)$ is for emphasis, and is not strictly necessary since the $p_i$ contains the information about $i$.} the eigenvectors of  $D^a(g_i)\otimes D^b(g_i)\otimes D^c(g_i)$ corresponding to the eigenvalue $-1$, then the common eigenvector(s) can be found by solving the following set of linear equations for the numerical coefficients $\beta ^{(i)}_{p_i}$:
\begin{align}
\sum _{p_1}\beta ^{(1)}_{p_1}\alpha ^{(1)}_{p_1}&=\sum _{p_2}\beta ^{(2)}_{p_2}\alpha ^{(2)}_{p_2} =\ldots =\sum _{p_{n-1}}\beta ^{(n-1)}_{p_{n-1}}\alpha ^{(n-1)}_{p_{n-1}}
\end{align}
$p_i$ counts the number of eigenvectors of $D^a(g_i)\otimes D^b(g_i)\otimes D^c(g_i)$ that correspond to eigenvalue $-1$. Note that the common eigenvector exists only if there exists a non-trivial solution for $\beta ^{(i)}_{p_i}$.

We emphasize once again that the eigenvectors $\alpha ^{(i)}_{p_i}$ can be written down fairly easily (without the need of a computer for low $n$, for example). So, we need to solve a set of linear equations that are less in number as compared to that of last section.

An alternate (and probably efficient in some cases) way of finding the common eigenvector for the set of matrices $A_i\equiv D^a(g_i)\otimes D^b(g_i)\otimes D^c(g_i)$ is as follows. We will use a variation of this approach to fix the Young patterns that appear at level 5. Suppose $x$ be the common eigenvector corresponding to the $-1$ eigenvalue to all the matrices $A_i$. This gives us:
\begin{align}
A_1x=A_2x=A_3x=\ldots =A_{n-1}x=-x
\end{align}
Adding all these equations gives us:
\begin{align}
\label{sum eigenvalue eqn}
Ax\equiv \left(A_1+A_2+A_3+\ldots +A_{n-1}\right)x&=-(n-1)x
\end{align}
This means that (some of) the eigenvectors of this eigenvalue equation are the common eigenvectors $x$. Note that we need to solve only a single eigenvalue equation in this case. %But, this way of finding the common eigenvectors necessarily needs a computer even for levels as low as $n=3$. 

One advantage of this method is that we have a simple way to get a bound on the number of common eigenvectors of the matrices $A_i$. We can do this as follows. Starting from \eqref{sum eigenvalue eqn}, we note that a non-zero eigenvector can exist only if:
\begin{align}
\text{det}\left(A+(n-1)I\right)&=0
\end{align} 
If this determinant is not zero, then $x$ has to be zero and thus there is no common eigenvector to the matrices $A_i$ corresponding to eigenvalue $-1$. Also, the number of common eigenvectors to all $A_i$ is less than or equal to the number of zero eigenvalues of the matrix $\left(A+(n-1)I\right)$. In terms of rank of the matrix, the last statement implies the following:
\begin{align}
\# \text{ of common eigenvectors}& \le \text{Order}\left(A+(n-1)I\right)-\text{Rank}\left(A+(n-1)I\right)
\end{align} 
This is a useful relation when checking for common eigenvectors using Mathematica.

 % \textbf{needs to be checked}.

Are there any other diagnostics (i.e., the ones that do not involve calculating determinants or eigenvalues) to find whether the matrices $A_i$ have common eigenvectors?  There is another way which is the most efficient while using a computer. We start by observing that $S_n$ is generated by only two elements- (12) and (12\ldots $n$). Our goal now is to find a common eigenvector between $D^a(12)\otimes D^b(12)\otimes D^c(12)$ and $D^a(12\ldots n)\otimes D^b(12\ldots n)\otimes D^c(12\ldots n)$ corresponding to eigenvalues $-1$ and $(-1)^{n-1}$ respectively. The technology developed above, of counting the rank and order, can be applied here as well. But instead of dealing with $(n-1)$ matrices as in the previous case, we deal here with only two\footnote{There is a simple way to check whether two matrices share a common eigenvector. If two matrices $B$ and $C$ share a common eigenvector, then we need to have $\text{det}[B,C]=0$.} matrices for any $S_n$. More precisely, the bound on the common eigenvectors is as follows:%\textbf{(CHECK THIS)}
\begin{align}
\# ~\text{of} &\text{common eigenvectors}\leq  \nonumber \\
&\text{Order}\left(D^a(12)\otimes D^b(12) \otimes D^c(12)+(-1)^nD^a(12\ldots n)\otimes D^b(12\ldots n) \otimes D^c(12\ldots n)+2 I_{d_ad_bd_c}\right)\nonumber \\
-&\text{Rank}\left(D^a(12)\otimes D^b(12) \otimes D^c(12)+(-1)^nD^a(12\ldots n)\otimes D^b(12\ldots n) \otimes D^c(12\ldots n)+2 I_{d_ad_bd_c}\right)
\end{align}
where $d_{a,b,c}$ are dimensions of the irreps $a,b,c$. Even though this method is strictly only an upper bound on the number of common eigenvectors, we found that at level 5 (which is the maximum level up to which we have done explicit calculations), the bounds are saturated. The allowed Young patterns \footnote{Remember that the Young patterns capture the representation and the standard Young tableaux list the basis elements.} for levels 4 and 5 are presented in appendices. The results are quite non-trivial, especially for the level 5 case, and we do not believe they can be obtained without the formalism arising from our main equation \eqref{main equation} in Section 1.

We conclude this section by giving a couple of examples on bounding the number of common eigenvectors at level 5. At level 5, we need to find common eigenvector(s) between $D^a(12)\otimes D^b(12)\otimes D^c(12)$ and $D^a(12345)\otimes D^b(12345)\otimes D^c(12345)$ corresponding to eigenvalues $-1$ and $1$ respectively. Note that the matrices $D^r(12)$ (and more generally the matrices $D^r(i,i+1)$) can be constructed easily for any representation $r\in S_5$ following the appendix \ref{yamanouchi basis} and the matrices $D^r(12345)$ can be obtained as follows:
\begin{align}
D^r(12345)&=D^r(12).D^r(23).D^r(34).D^r(45)
\end{align}
Let us start by choosing all of $a,b,c$ to be the sign representation of $S_5$ i.e., 
\begin{figure}[htbp!]
	\centering
	\includegraphics[trim={4cm 24.2cm 3cm 2.5cm},clip,scale=1]{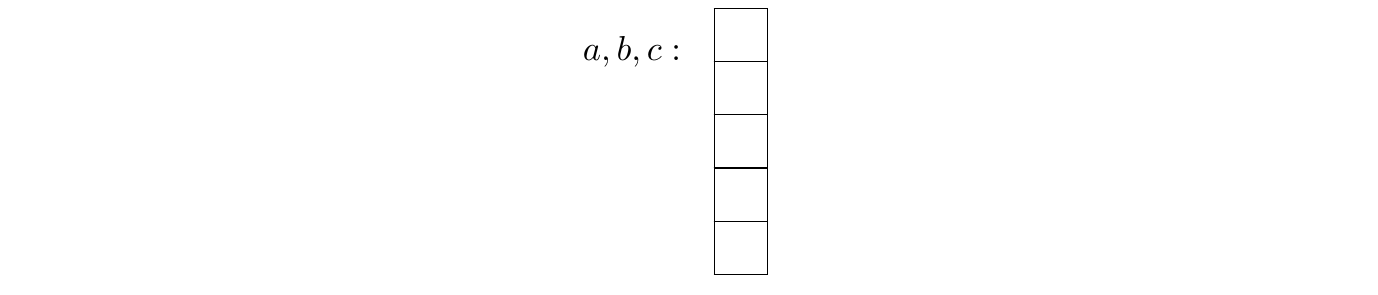}
\end{figure}\\
%\begin{align}
%a,b,c: ~~\begin{ytableau}
%~ \\ ~ \\ ~ \\ ~ \\ ~\\
%\end{ytableau}
%\end{align}
Sign representation is a one dimensional representation and in this representation, we have $D^a(12)=D^b(12)=D^c(12)=-1$ and $D^a(12345)=D^b(12345)=D^c(12345)=1$. Hence, there may be a common eigenvector in this case and we find that all of $a,b,c$ being the sign representation is an antisymmetric state with multiplicity 1 \footnote{By multiplicity, we mean the number of times a set of Young patterns appear in the set of antisymmetric states at a particular level.}.

Next, we move on to a slightly complicated example. Let us take: 
\begin{figure}[!htbp]
	\centering
	\includegraphics[trim={4cm 24.7cm 3cm 2.5cm},clip,scale=1]{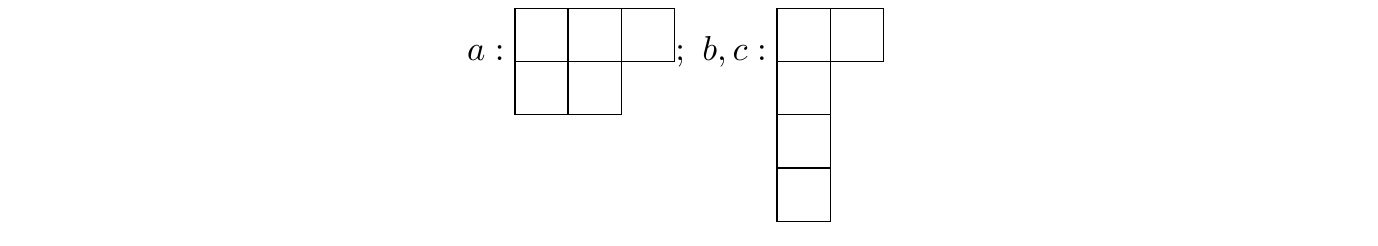}
\end{figure}
%\begin{align}
%a:\begin{ytableau}~&~&~ \\ ~&~ 
%\end{ytableau};~
%b,c:\begin{ytableau}~&~ \\ ~ \\ ~ \\ ~ 
%\end{ytableau}
%\end{align}
\newpage
``$a$'' is a 5 dimensional representation and ``$b,c$'' are 4 dimensional representations. Following appendix \ref{yamanouchi basis}, we can write the following representation matrices:
\begin{align}
D^a(12)&=\diag (1,1,1,-1,-1); \hspace{32mm} D^{b,c}(12)=\diag (-1,-1,-1,1)\nonumber \\
D^a(12345)&=\begin{pmatrix}
-\frac{1}{3} & -\frac{\sqrt{2}}{3} &\sqrt{\frac{2}{3}} & 0 & 0 \\ -\frac{\sqrt{2}}{3} & \frac{1}{12} & -\frac{1}{\sqrt{48}} & -\frac{\sqrt{3}}{4} & \frac{3}{4} \\ 0 & -\frac{\sqrt{3}}{4} & -\frac{1}{4} & -\frac{3}{4} & -\frac{\sqrt{3}}{4} \\ -\sqrt{\frac{2}{3}} & \frac{1}{\sqrt{48}} & -\frac{1}{4} & \frac{1}{4} & -\frac{\sqrt{3}}{4} \\ 0 & -\frac{3}{4} & -\frac{\sqrt{3}}{4} & \frac{\sqrt{3}}{4} & \frac{1}{4}
\end{pmatrix}; ~~
D^{b,c}(12345)=\begin{pmatrix}
-\frac{1}{4} & -\frac{\sqrt{15}}{4} &0 &0 \\ \sqrt{\frac{5}{48}} &-\frac{1}{12} & -\frac{2\sqrt{2}}{3} &0 \\ -\sqrt{\frac{5}{24}} &\frac{1}{\sqrt{72}} & -\frac{1}{6} &-\frac{\sqrt{3}}{2}  \\ \sqrt{\frac{5}{8}} &-\frac{1}{\sqrt{24}} & \frac{1}{\sqrt{12}} &-\frac{1}{2}
\end{pmatrix}
\end{align}
By explicit computation, we see that 
\begin{align}
&\text{Order}\left(D^a(12)\otimes D^b(12) \otimes D^c(12)-D^a(12345)\otimes D^b(12345) \otimes D^c(12345)+2 I_{80}\right)\nonumber \\
&~~~~=\text{Rank}\left(D^a(12)\otimes D^b(12) \otimes D^c(12)-D^a(12345)\otimes D^b(12345) \otimes D^c(12345)+2 I_{80}\right)
\end{align}
and hence there are no common eigenvectors between $D^a(12)\otimes D^b(12)\otimes D^c(12)$ and $D^a(12345)\otimes D^b(12345)\otimes D^c(12345)$ corresponding to eigenvalues $-1$ and $1$. Hence, this choice of $a,b,c$ does \textit{not} lead to an antisymmetric state. In a similar way, we can check for all the possible choices of $a,b,c$ to find the possible young patterns that can be antisymmetric states. The complete set of young patterns that are antisymmetric at levels 4 and 5 is listed in the appendices.  

A sanity check of our results is to compare the sum of the number states in all these representations together, with the total number of fully anti-symmetric states at that level, where we treat the quantum numbers as belonging to specific groups. We explain this in the next section.

%But there is no simple way to write down the matrix $D(12\ldots n)$ or its eigenvalues or eigenvectors. Hence, the use of computer is necessary even for lower levels.

\section{Counting States as a Sanity Check}

In this section, we present a way to verify the antisymmetric states we found are correct by counting the dimensions of the irreps of the groups to which the quantum numbers belong. Consider the fermions  of the form $\psi ^{ijk}$ carrying three quantum numbers corresponding to $SU(n_1)_i\times SU(n_2)_j\times SU(n_3)_k$. Denoting the indices $ijk\equiv I$, we can see that any state with multiple fermions is in completely antisymmetric representation of $SU(n_1  n_2 n_3)$ i.e., a state at level $n$ is given by the following representation of $SU(n_1 n_2 n_3)$:
\begin{figure}[h!]
	\centering
	\includegraphics[trim={4cm 24.7cm 3cm 2.5cm},clip,scale=1]{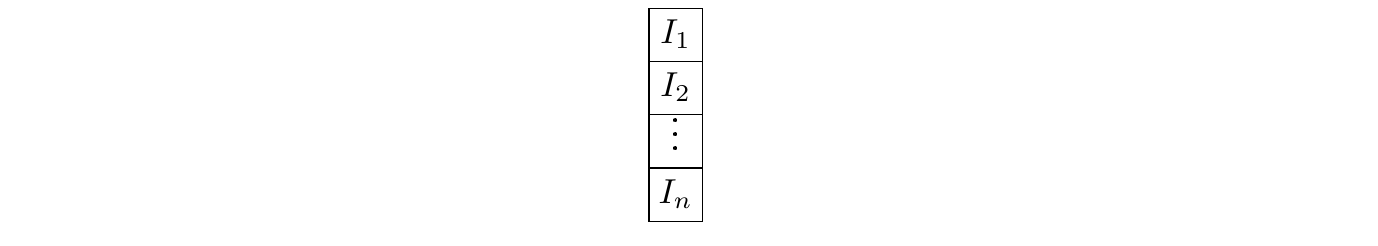}
\end{figure}\\
%\begin{align}
%\begin{ytableau}I_1\\I_2\\\vdots \\I_n \end{ytableau}
%\end{align}  
The number of rows here is equal to the level we are considering, namely $n$. 
The dimension of this representation for $SU(n_1  n_2 n_3)$ is trivial to calculate using hook rule (say). This dimension should be exactly equal to the sums of dimensions of various antisymmetric representations (aka Young patterns) of $SU(n_1)_i\times SU(n_2)_j\times SU(n_3)_k$ we find by solving our main equation \eqref{main equation}: in particular, this should apply for the level 4 and level 5 cases we have listed in the appendix. This provides a non-trivial check, and all the representations we find for two and three-index cases do pass this check. 

As an illustration, we present here how the counting works for the level 4 in the 3-slot case. For simplicity, we choose $n_1=n_2=n_3=n$. That is, we explicitly show that the dimensions of the completely antisymmetric representation of $SU(n^3)$ at level 4 is equal to the $SU(n)_i\times SU(n)_j\times SU(n)_k$ representations listed in appendix \ref{level4} . Using hook rule,  the dimension of completely antisymmetric representation of $SU(n^3)$ at level-4 is given by ${n^3}\choose{4}$. The sum of dimensions of the representations in the same order as in appendix \ref{level4} are given as follows:
\begin{align}
\label{sum-level4}
&6\times d_{a_2}d_{a_1}d_{a_4}+6\times d_{a_2}d_{a_3}d_{a_4}+3\times d_{a_2}d_{a_2}d_{a_5}+3\times d_{a_4}d_{a_4}d_{a_5}+3\times d_{a_1}d_{a_1}d_{a_5}+3\times d_{a_3}d_{a_3}d_{a_5}+3\times d_{a_3}d_{a_3}d_{a_1}\nonumber \\
+&3\times d_{a_3}d_{a_2}d_{a_2}+3\times d_{a_3}d_{a_4}d_{a_4}+3\times d_{a_2}d_{a_2}d_{a_4}+3\times d_{a_4}d_{a_4}d_{a_2}+ d_{a_2}d_{a_2}d_{a_2}+d_{a_4}d_{a_4}d_{a_4}+d_{a_3}d_{a_3}d_{a_3}+d_{a_5}d_{a_5}d_{a_5}
\end{align}
where $d_{a_1}\ldots d_{a_5}$ are respectively the dimensions of the  irreps of $SU(n)$ at level-4:
\begin{figure}[h!]
	\centering
	\includegraphics[trim={4cm 24.7cm 3cm 2.5cm},clip,scale=1]{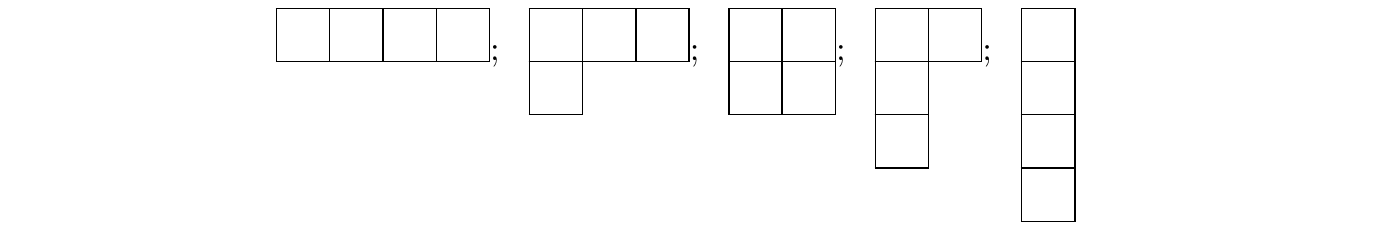}
\end{figure}\\
%\begin{align}
%\begin{ytableau}
%~&~&~&~
%\end{ytableau};~~\begin{ytableau}
%~&~&~\\ ~
%\end{ytableau};~~\begin{ytableau}
%~&~\\ ~&
%\end{ytableau};~~\begin{ytableau}
%~&~\\~\\ ~
%\end{ytableau};~~\begin{ytableau}\\~\\~ \\~ \end{ytableau}
%\end{align}
From the hook rule, the dimensions of these tableau can be computed to be:
\begin{align}
d_{a_1}&=\frac{n(n+1)(n+2)(n+3)}{24}; ~~ d_{a_2}=\frac{n(n+1)(n+2)(n-1)}{8}; ~~ d_{a_3}=\frac{n^2(n+1)(n-1)}{12}\nonumber \\
 d_{a_4}&=\frac{n(n-1)(n-2)(n+1)}{8}; ~~d_{a_5}=\frac{n(n-1)(n-2)(n-3)}{24}
\end{align}
Substituting these values, we get the above sum in \eqref{sum-level4} to be $n^3\choose 4 $ which matches with the antisymmetric representation of $SU(n^3)$ at level 4 as expected. Note that the counting works even for different $n_1,n_2,n_3$. 

Further, we have verified that the counting works for the 3-slot case up to level 5 for antisymmetric representations and for the 2-slot case up to level 6 for both symmetric and antisymmetric representations.

\section*{Acknowledgments}

We thank B. Anathanarayan for comments on baryon wave functions and Avinash Raju, Dario Rosa and Sambuddha Sanyal for related collaborations.

\appendix

\section{Symmetric Groups and Young Tableaux: Mini Review}
The Symmetric group $ S_{n} $ represents all the permutations that can be performed on a finite set of $ n $ symbols \cite{Coleman}. It is composed of cycles, of length $ l\leq n $, which preserve $ (n-l) $ elements in the set invariant. Two different cycles of the same length are conjugate to each other. Hence, the length of the cycles defines the conjugacy classes of the symmetric group. We can take a set of integers $ \lambda_{i} $ such that  $\sum_{i}\lambda_{i} = n $, which is the partition of $ n $, which defines the conjugacy class of that particular set of cycles.

The symmetric group, as mentioned above, is characterized by the cycles. Thus, the cycles can be understood as the generating set of the group. The group $ S_{n} $ can be generated by different set of cycles. For $ n\geq 3 $, any cycle of length greater than two can be generated by an appropriate product 2-cycles (also called transpositions). The 2-cycles are represented as $ (i_{j}i_{j+1}) $, and we have
\bea
(i_{1}i_{2}\dots i_{k}) = (i_{1}i_{2})(i_{2}i_{3})\dots (i_{k-1}i_{k}).
\eea
We can generate all cycles using this method, which in turn means that we can generate the entire group $ S_{n} $. In this method, the generating set is made up of $ {}^{n}C_{2} = n(n-1)/2 $ transpositions. 

We could also generate all the cycles using $ n-1 $ transpositions chosen in two different ways, for $ n\geq 3 $:
\bea
&&(12), (13), \dots ,(1n)\\
&&(12), (23), \dots , (n-1\, n).
\eea
The generating set can also be constructed out of just two elements in the following way, again for $ n\geq 3 $:
\bea
&&(12),(123\dots n)\\
&&(12),(23\dots n).
\eea

As mentioned above, the symmetric group is characterized entirely by the cycles, which in turn are unique only upto their conjugacy class. The conjugacy classes are defined only by the lengths of the cycles it contains, and all such classes are given by the partition of $ n $. This is where Young tableaux enters the picture. Young tableaux is a nice graphic way of representing partitions. Two crucial facts are:
\begin{itemize}
\item The irreducible representations of the group $S_n$ are labelled by the Young patterns of $n$. 
\item The basis elements of a given irrep (aka Young pattern) are labelled by the standard Young tableaux corresponding to that pattern.
\end{itemize}

We define and discuss the words in the above paragraph now, in some detail. From the partitions of $ n $, the first thing one can define is a \emph{Young pattern} $ [\lambda] $ in the following way
\bea
&&[\lambda] = [\lambda_{1},\lambda_{2},\dots, \lambda_{m}], \ \ \ \text{such that} \sum_{k=1}^{m}\lambda_{k} = n ,\\
&&\lambda_{1}\geq \lambda_{2}\geq \dots\lambda_{m} \geq 0.
\eea
The Young pattern $ [\lambda] $ can be graphically represented by a set of left justified boxes, where the $ i $-th row contains $ \lambda_{i} $ boxes. The last line in the above relation will then enforce the condition that the number of boxes in any row is atleast as many as the number of boxes in the row just below it, and that the number of boxes in any column is atleast as many the number of boxes in the column just to the right of it. For example, the Young pattern $ [\lambda] = [4,2,2,1] $ is graphically represented as 
\begin{figure}[h!]
	\centering
	\includegraphics[trim={4cm 24.5cm 3cm 2.5cm},clip,scale=1]{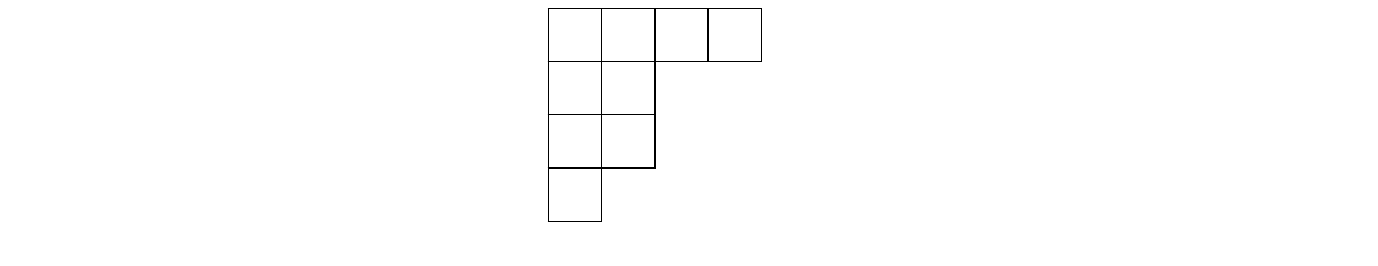}
\end{figure}\\
%\bea
%\begin{ytableau}
%~	&~&~&~\\ ~&~\\ ~&~\\ ~
%\end{ytableau}
%\eea
It is not allowed to have a Young pattern in the form $ [\lambda] = [4,2,1,2] $, although it is a partition of 9, as it will not satisfy the rules.

A Young pattern with the each of the boxes being assigned a number $ 1,2,\dots,n $ is called a \emph{Young tableaux}. It is evident that there will be $ n! $ Young tableau for any given Young pattern. If the numbers assigned to the boxes obeys the rule: the values are increase as we go left to right along every row and top to bottom along every column, then it is called a \emph{standard Young tableaux}. Consider the following Young tableau for example\\
\begin{figure}[h!]
	\centering
	\includegraphics[trim={4cm 25.5cm 3cm 2.5cm},clip,scale=1]{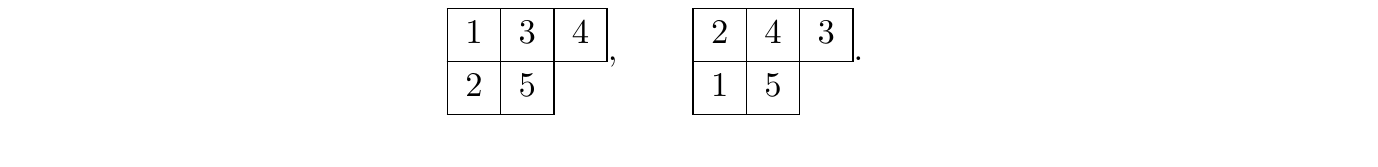}
\end{figure}\\
%\bea
%\begin{ytableau}
%	1& 3&4\\ 2&5
%\end{ytableau} ,\qquad \begin{ytableau}
%2& 4 &3 \\ 1& 5  
%\end{ytableau}.
%\eea
Although both are valid Young tableau, only the first one qualifies to be a standard Young tableaux. The number of standard Young tableau for a given Young pattern can be easily determined by the hook length formula. The hook length $ h_{ij} $ of a box at $ i $-th row $ j $-th column is given by the sum of the number of boxes to its right in that column and the number of boxes below it in that column added to 1. The number of standard tableau in the Young pattern is then given by
\bea
d_{[\lambda]}(S_{n} ) = n!\prod_{ij}\dfrac{1}{h_{ij}}.
\eea 

\newpage
For example, the hook lengths and the corresponding number of standard tableau for the Young pattern $ [\lambda] = [3,2,2,1] $ is 
\begin{figure}[h!]
	\centering
	\includegraphics[trim={4cm 24.5cm 3cm 2.5cm},clip,scale=1]{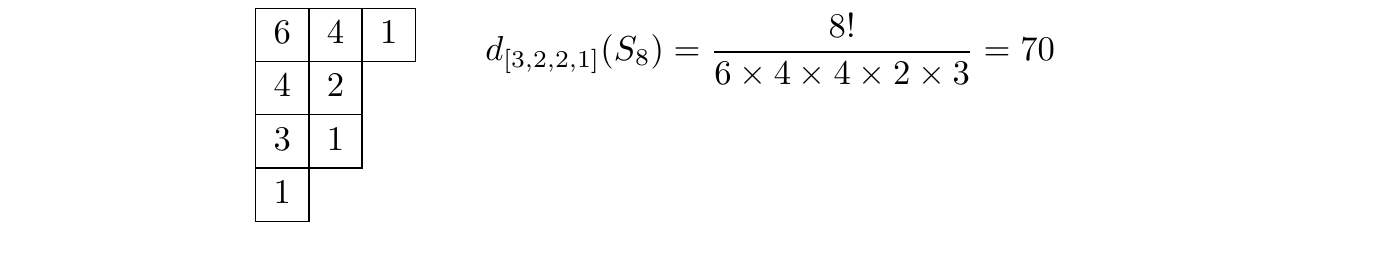}
\end{figure}\\
%\bea
%\begin{ytableau}
%	6&4&1\\ 4&2\\ 3&1\\ 1
%\end{ytableau}\qquad d_{[3,2,2,1]}(S_{8}) = \dfrac{8!}{6 \times 4\times 4 \times2 \times 3} = 70
%\eea
Consider two permutation groups on a finite set of $ n+m $ elements, first one $ S_{n} $ acting on the first $ n $ objects and the second one $ S_{m} $ acting on the last $ m $ objects. The two permutations act on two different subsets, which means they commute. This particular permutation action can be understood as the $ S_{n}\otimes S_{m} $ subgroup of the group $ S_{n+m} $. We can represent the subgroup as a sum of the irreducible representations of the group $ S_{n+m} $, using the \emph{Littlewood-Richardson Rule}. That is, a Young pattern $ [\lambda] $ of $ S_{n} $ and $ [\xi] $ of $ S_{m} $ can be combined into a sum over representations $ [\mu] $ of $ S_{n+m} $ as
\bea
[\lambda]\otimes [\xi] = \bigoplus_{[\mu]} C^{[\mu]}_{[\lambda][\xi]}\, [\mu],
\eea
where $ C^{[\mu]}_{[\lambda][\xi]} $ is the multiplicity with which $ [\mu] $ occurs. 

For combining representations, start by labelling the $ i $-th row of $ [\xi] $ with the digit $ i $. Then attach these boxes of $ \xi $ row by row, ordered top to bottom, to the Young pattern $ [\lambda] $ following the rules:
\begin{itemize}
	\item Each time a row of $ [\xi] $ is attached to $ [\lambda] $, the resultant pattern is a valid Young pattern.
	\item None of the columns in the resultant pattern has a repeated digit.
	\item After all the boxes in $ [\xi] $ have been attached to $ [\lambda] $, the number of boxes with a smaller digit is never less than the number of boxes with a larger digit when read right to left, in each row.
\end{itemize}

As an example, consider combining the Young patterns $ [2,1]\otimes [2,1] $. First we label the rows in the second Young pattern
\begin{figure}[h!]
	\centering
	\includegraphics[trim={4cm 25.7cm 3cm 2.5cm},clip,scale=1]{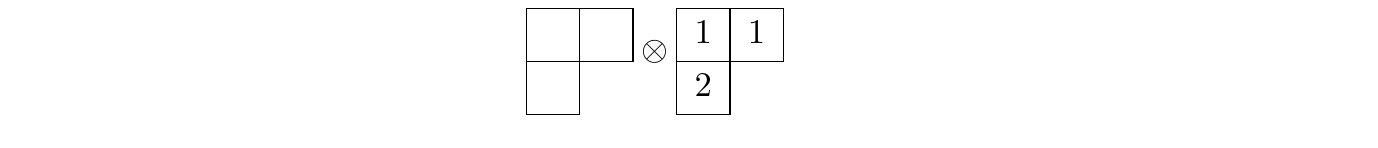}
\end{figure}\\
%\bea
%\begin{ytableau}
%	~&~\\~
%\end{ytableau} \otimes \begin{ytableau}
%1&1\\2
%\end{ytableau}
%\eea
The first row is attached to the Young pattern $ [\lambda] $ to get
\begin{figure}[h!]
	\centering
	\includegraphics[trim={4cm 25.2cm 3cm 2.5cm},clip,scale=1]{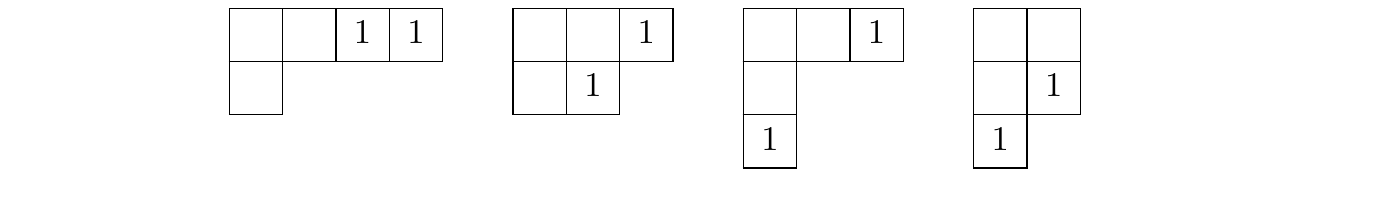}
\end{figure}\\
%\bea\label{firstrow}
%\begin{ytableau}
%	~&~&1&1\\~
%\end{ytableau}\qquad \begin{ytableau}
%~&~&1\\~&1
%\end{ytableau} \qquad \begin{ytableau}
%~&~&1\\~\\1
%\end{ytableau} \qquad \begin{ytableau}
%~&~\\~&1\\1
%\end{ytableau}
%\eea
Then attach the second row, which by the rules, gives the allowed patterns:
\begin{figure}[h!]
	\centering
	\includegraphics[trim={4cm 23cm 3cm 2.5cm},clip,scale=0.92]{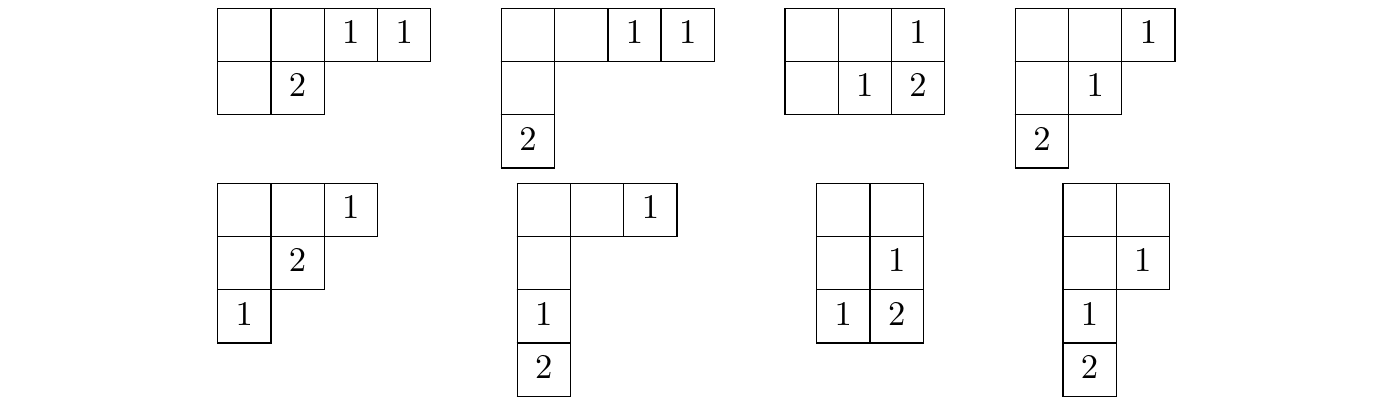}
\end{figure}\\
%\bea
%&&\begin{ytableau}
%	~&~&1&1\\~&2
%\end{ytableau}\qquad \begin{ytableau}
%~&~&1&1\\~\\ 2
%\end{ytableau}\qquad \begin{ytableau}
%~&~&1\\~&1 &2
%\end{ytableau} \qquad\begin{ytableau}
%~&~&1\\~&1 \\2
%\end{ytableau} \non\\
%&&\begin{ytableau}
%~&~&1\\~&2\\1
%\end{ytableau} \qquad\qquad \begin{ytableau}
%~&~&1\\~\\1\\2
%\end{ytableau}  \qquad\qquad \begin{ytableau}
%~&~\\~&1\\1&2
%\end{ytableau} \qquad\qquad \begin{ytableau}
%~&~\\~&1\\1\\2
%\end{ytableau}
%\eea
\newpage
For clarity, the rules do not allow for a box with digit 2 attached to the first row and hence such Young patterns are discarded here. The multiplicities are given by
\bea
C^{[\mu]}_{[\lambda][\xi]} = \begin{cases}
	2,\qquad \text{for }[\mu] = [3,2,1]\\
	1,\qquad \text{otherwise}.
\end{cases}
\eea

\section{Young-Yamanouchi Basis}\label{yamanouchi basis}

Young tableaux are a useful way to describe various irreducible representations of the symmetry group and also of unitary and orthogonal groups. We define standard Young tableaux as a tableaux where the numbers in a row increase from left to right and the numbers in a column increase from top to bottom. Also, we use a convention that a state corresponding to a standard Young tableaux is obtained by first symmetrizing along the rows and then antisymmetrizing along the columns. For example,
\begin{figure}[h!]
	\centering
	\includegraphics[trim={4cm 26cm 3cm 2.5cm},clip,scale=1]{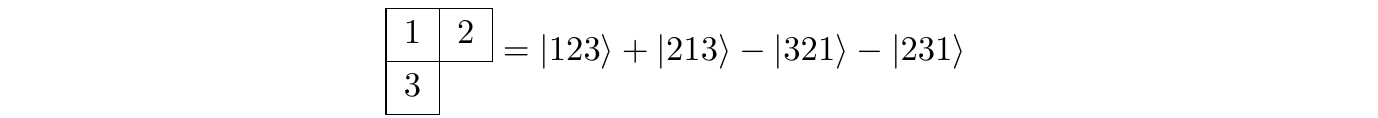}
\end{figure}\\
%\begin{align}
%\begin{ytableau}1&2\\3\end{ytableau}&=|123\rangle +|213\rangle -|321\rangle -|231\rangle
%\end{align}   
Further, it can be shown that the standard Young tableaux form a basis and hence any state can be expanded in terms of standard tableaux. This fact is important to write down the equation \eqref{main equation}.

In the Young-Yamanouchi basis, we represent all the standard Young tableaux of a particular representation by column matrices where only one of the quantities is $1$ and the others are zero. Now, the  matrix elements of the form $\langle a'|D(g)|a\rangle $ are constructed where $D(g)$ is the matrix corresponding to the 2-cycle $g$ in the specific representation we are dealing with. Note that we need to only find the matrices corresponding to 2-cycles of the form $(i,i+1)$ as all the entire permutation group can be generated by these transpositions.

The matrix elements corresponding to $D(i,i+1)$ are quite simple and are given as follows. If the standard Young tableaux $|a\rangle $ and $|a'\rangle $ are such that they can be obtained by exchanging $i$ and $(i+1)$, then we have:
\begin{align}
D(i,i+1)|a\rangle =-\rho _{(i,i+1)}|a\rangle +\sqrt{1-\rho _{(i,i+1)}^2} ~|a'\rangle \\
D(i,i+1)|a'\rangle =+\sqrt{1-\rho _{(i,i+1)}^2} ~|a\rangle +\rho _{(i,i+1)}|a'\rangle
\end{align}
where $\rho _{(i,i+1)}$ is the inverse of the distance between $i$ and $(i+1)$ i.e., the number of steps taken from $i$ to reach $(i+1)$. When the steps are counted left or down, we take them to be positive distance. Right or upward steps contribute to negative distance. If $|a\rangle $ and $|a'\rangle $ are two standard tableaux such that they are not obtained by the exchange of $i$ and $(i+1)$, then $\langle a'|D(i,i+1)|a\rangle =0$.

If $i$ and $(i+1)$ are in the same row and adjacent to each other in a certain standard tableaux then $\rho _{(i,i+1)}=-1$ and we put $1$ in the corresponding position and the rest of the entries in that column and row are zeroes. Similarly, for $i$ and $(i+1)$ in the same column, we put $-1$ in the corresponding position and the rest of the entries are zeroes.  

As an example, consider the following standard tableaux at level 4:
\begin{figure}[h!]
	\centering
	\includegraphics[trim={4cm 26cm 3cm 2.5cm},clip,scale=1]{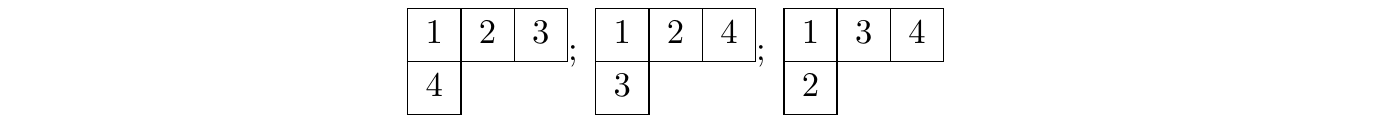}
\end{figure}\\
%\begin{align}
%\begin{ytableau}1&2&3\\4\end{ytableau}; ~\begin{ytableau}1&2&4\\3\end{ytableau}; ~\begin{ytableau}1&3&4\\2\end{ytableau}
%\end{align}
We assign the following column matrices to the above tableaux as follows:
\begin{figure}[h!]
	\centering
	\includegraphics[trim={4cm 25.7cm 3cm 2.5cm},clip,scale=1]{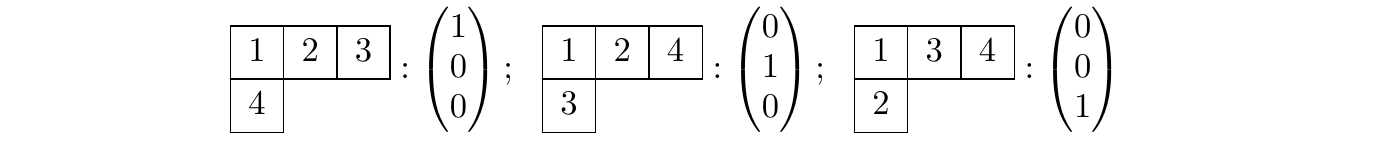}
\end{figure}\\
%\begin{align}
%\begin{ytableau}1&2&3\\4\end{ytableau}:\begin{pmatrix}1\\0\\0\end{pmatrix}; ~~\begin{ytableau}1&2&4\\3\end{ytableau}:\begin{pmatrix}0\\1\\0\end{pmatrix}; ~~\begin{ytableau}1&3&4\\2\end{ytableau}:\begin{pmatrix}0\\0\\1\end{pmatrix}
%\end{align} 
Following the above rules, we get the matrices corresponding to $(12)$, $(23)$ and $(34)$ as:
\begin{align}
D(12)&=\begin{pmatrix}
1&0&0\\0&1&0\\0&0&-1
\end{pmatrix};~~
D(23)=\begin{pmatrix}
1&0&0\\0&-\frac{1}{2}&\frac{\sqrt{3}}{2}\\0&\frac{\sqrt{3}}{2}&\frac{1}{2}
\end{pmatrix};~~
D(34)=\begin{pmatrix}
-\frac{1}{3}&\frac{\sqrt{8}}{3}&0\\\frac{\sqrt{8}}{3}&\frac{1}{3}&0\\0&0&1
\end{pmatrix}
\end{align}

\section{Two Slots: Symmetric and Anti-Symmetric states}

\begin{comment}
\subsection{Level 2}
Symmetric States
\bea
\left( \begin{ytableau}
	~&~
\end{ytableau}\ \ , \begin{ytableau}
~&~
\end{ytableau}\right),\qquad 
\left( \begin{ytableau}
	~\\~
\end{ytableau}\ \ , \begin{ytableau}
~\\~
\end{ytableau}\right)
\eea
Anti-Symmetric states
\bea
\left( \begin{ytableau}
	~&~
\end{ytableau}\ \ , \begin{ytableau}
~\\~
\end{ytableau}\right),\qquad 
\left( \begin{ytableau}
	~\\~
\end{ytableau}\ \ , \begin{ytableau}
~&~
\end{ytableau}\right)
\eea
\subsection{Level 3}
Symmetric states
\bea
\left(\begin{ytableau}
~&~&~
\end{ytableau}\ , \begin{ytableau}
~&~&~
\end{ytableau} \right),\qquad\left(\ \begin{ytableau}
~&~\\~
\end{ytableau}\ , \begin{ytableau}
~&~\\~
\end{ytableau} \right),\qquad\left(\ \begin{ytableau}
~\\~\\~
\end{ytableau}\ , \begin{ytableau}
~\\~\\~
\end{ytableau} \ \right)
\eea

Anti-Symmetric states
\bea
\left(\begin{ytableau}
	~&~&~
\end{ytableau}\ , \begin{ytableau}
~\\~\\~
\end{ytableau} \right),\qquad\left(\ \begin{ytableau}
~&~\\~
\end{ytableau}\ , \begin{ytableau}
~&~\\~
\end{ytableau} \right),\qquad\left(\ \begin{ytableau}
~\\~\\~
\end{ytableau}\ , \begin{ytableau}
~&~&~
\end{ytableau} \ \right)
\eea
\end{comment}

The Young patterns of the symmetric and anti-symmetric states with two slots have a simple structure. Instead of explaining the structure with a thousand words, we will present the pictures of the corresponding patterns at levels 4 and 5, which have enough structure to illustrate the idea. It is easy to check that this structure holds at all levels, and that the total dimensionalities of each of these representations add up to the total number of symmetric and anti-symmetric states expected at each level.
\subsection{Level 4}
Symmetric states
\begin{figure}[h!]
	\centering
	\includegraphics[trim={4cm 22cm 3cm 2.5cm},clip,scale=1]{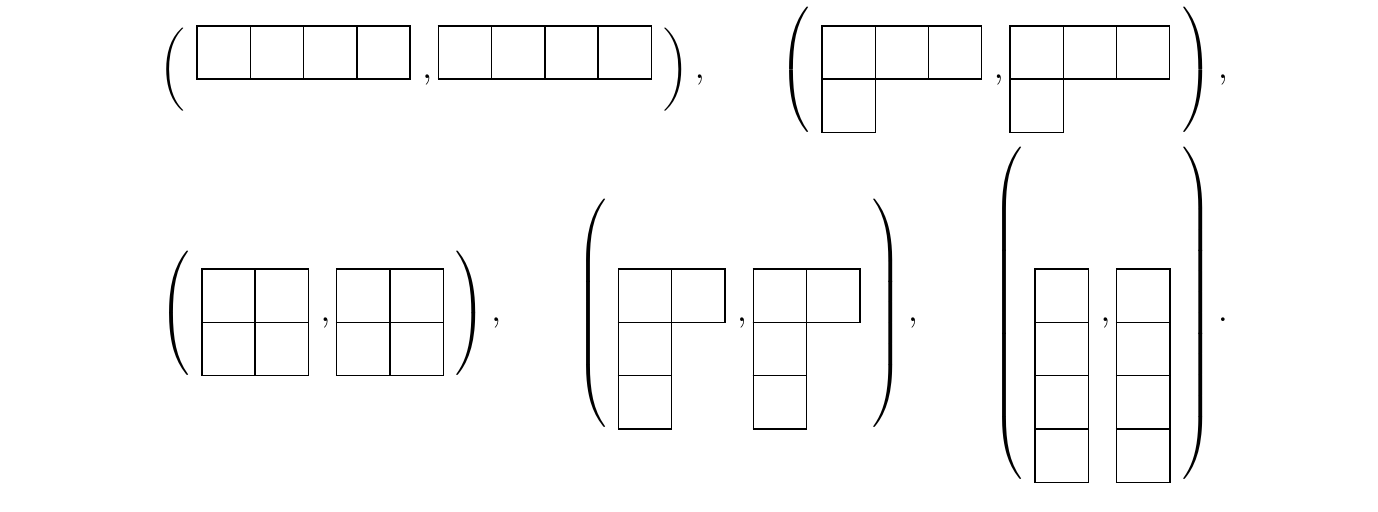}
\end{figure}\\
\begin{comment}
\bea
&&\left(\ \begin{ytableau}
~&~&~&~
\end{ytableau}\ ,\begin{ytableau}
~&~&~&~
\end{ytableau} \  \right),\qquad
\left(\ \begin{ytableau}
	~&~&~\\~
\end{ytableau}\ ,\begin{ytableau}
~&~&~\\~
\end{ytableau} \  \right),\non\\
&&\left(\ \begin{ytableau}
~&~\\~&~
\end{ytableau}\ ,\begin{ytableau}
~&~\\~&~
\end{ytableau} \  \right),\qquad\left(\ \begin{ytableau}
~&~\\~\\~
\end{ytableau}\ ,\begin{ytableau}
~&~\\~\\~
\end{ytableau} \  \right),\qquad\left(\ \begin{ytableau}
~\\~\\~\\~
\end{ytableau}\ ,\begin{ytableau}
~\\~\\~\\~
\end{ytableau} \  \right).
\eea
\end{comment}
Anti-Symmetric states
\begin{figure}[h!]
	\centering
	\includegraphics[trim={4cm 20cm 3cm 2.5cm},clip,scale=1]{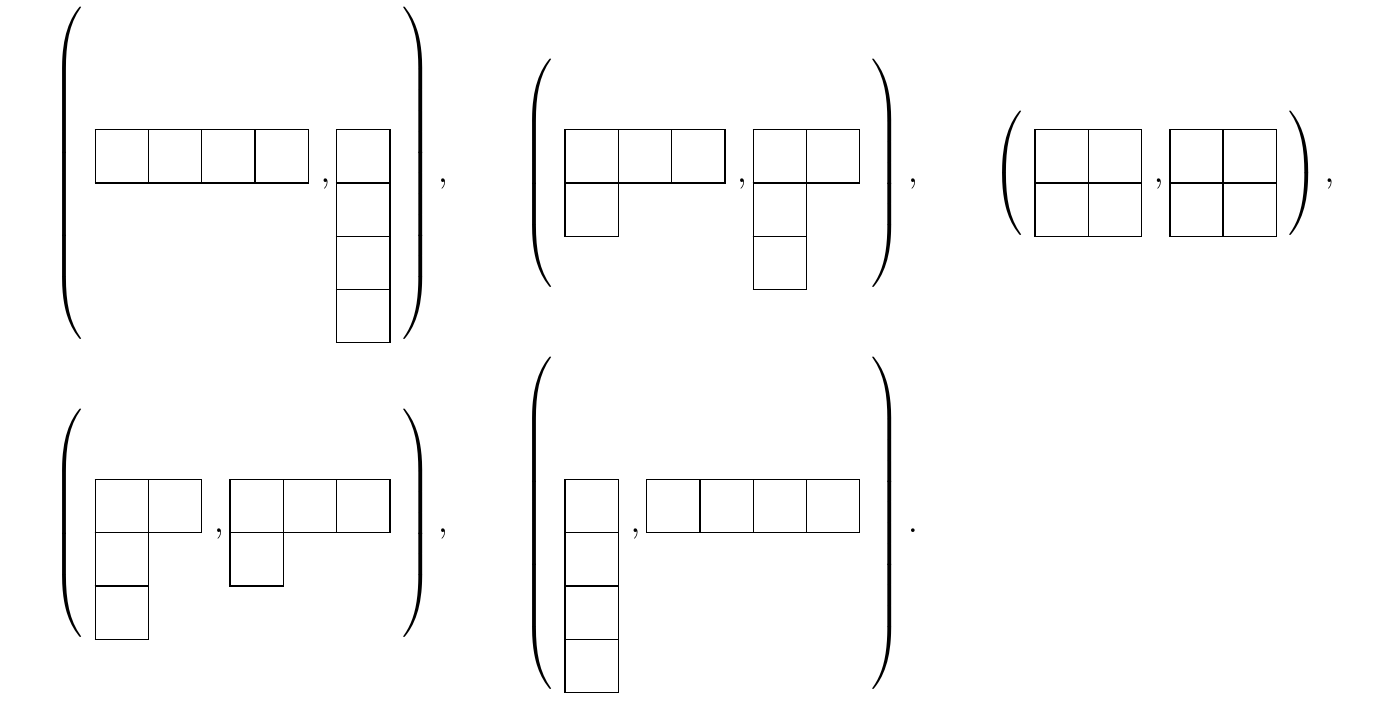}
\end{figure}\\
\begin{comment}
\bea
&&\left(\ \begin{ytableau}
	~&~&~&~
\end{ytableau}\ ,\begin{ytableau}
~\\~\\~\\~
\end{ytableau} \  \right),\qquad
\left(\ \begin{ytableau}
	~&~&~\\~
\end{ytableau}\ ,\begin{ytableau}
~&~\\~\\~
\end{ytableau} \  \right),\qquad \left(\ \begin{ytableau}
	~&~\\~&~
\end{ytableau}\ ,\begin{ytableau}
~&~\\~&~
\end{ytableau} \  \right),\non\\
&& \left(\ \begin{ytableau}
~&~\\~\\~
\end{ytableau}\ ,\begin{ytableau}
~&~&~\\~
\end{ytableau} \  \right),\qquad\left(\ \begin{ytableau}
~\\~\\~\\~
\end{ytableau}\ ,\begin{ytableau}
~&~&~&~
\end{ytableau} \  \right).
\eea
\end{comment}

\newpage
\subsection{Level 5}
Symmetric states
\begin{figure}[h!]
	\centering
	\includegraphics[trim={4cm 18.5cm 3cm 2.5cm},clip,scale=1]{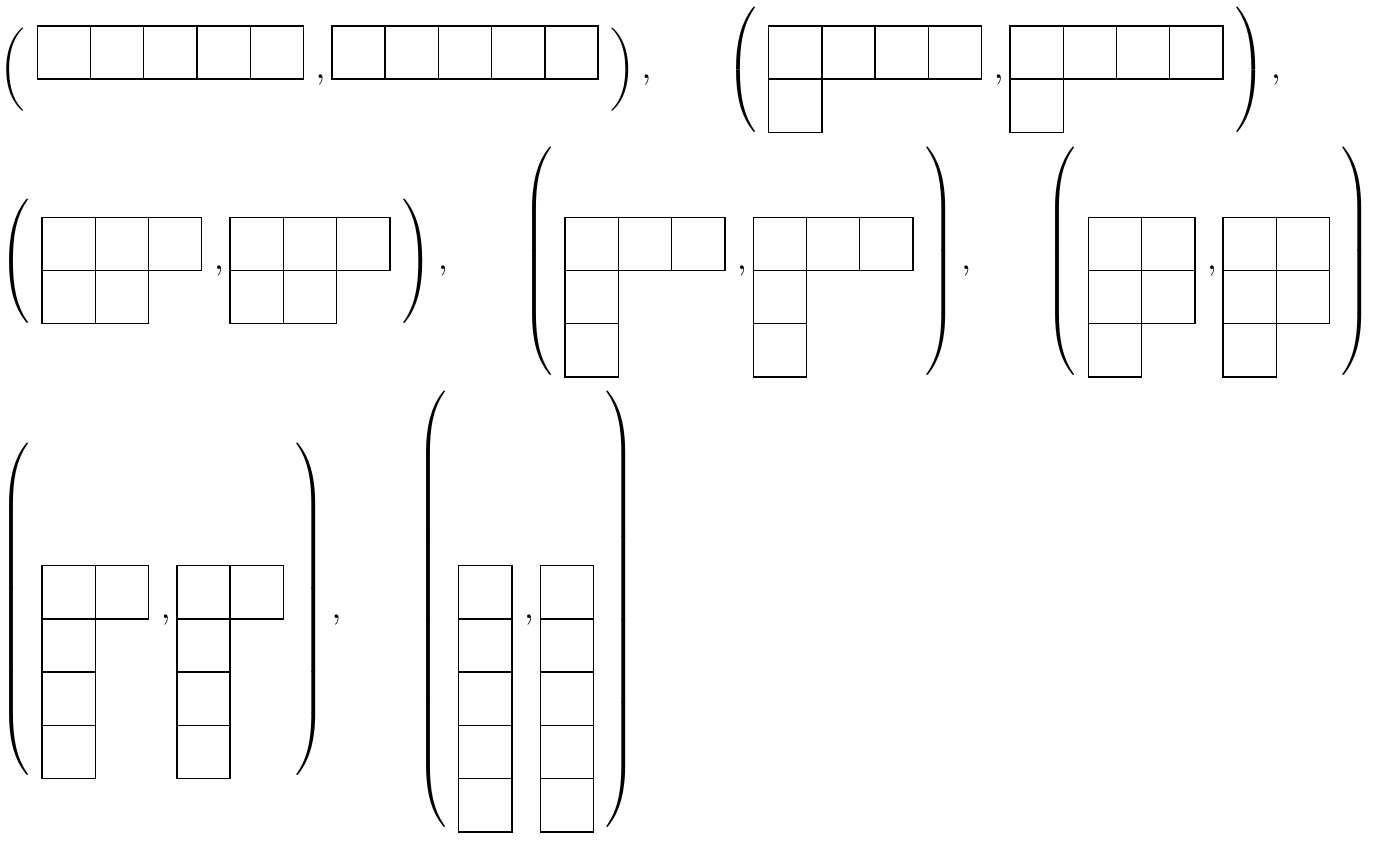}
\end{figure}\\
\begin{comment}
\bea
&&\left(\ \begin{ytableau}
	~&~&~&~&~
\end{ytableau}\ , \begin{ytableau}
~&~&~&~&~
\end{ytableau} \ \right),\qquad
\left(\ \begin{ytableau}
	~&~&~&~\\~
\end{ytableau}\ , \begin{ytableau}
~&~&~&~\\~
\end{ytableau} \ \right),\non\\
&&\left(\ \begin{ytableau}
	~&~&~\\~&~
\end{ytableau}\ , \begin{ytableau}
~&~&~\\~&~
\end{ytableau} \ \right),\qquad
\left(\ \begin{ytableau}
	~&~&~\\~\\~
\end{ytableau}\ , \begin{ytableau}
~&~&~\\~\\~
\end{ytableau} \ \right),\qquad
\left(\ \begin{ytableau}
	~&~\\~&~\\~
\end{ytableau}\ , \begin{ytableau}
~&~\\~&~\\~
\end{ytableau} \ \right),\non\\
&& \left(\ \begin{ytableau}
	~&~\\~\\~\\~
\end{ytableau}\ , \begin{ytableau}
~&~\\~\\~\\~
\end{ytableau} \ \right),\qquad \left(\ \begin{ytableau}
~\\~\\~\\~\\~
\end{ytableau}\ , \begin{ytableau}
~\\~\\~\\~\\~
\end{ytableau} \ \right)
\eea
\end{comment}
Anti-Symmetric states
\begin{figure}[h!]
	\centering
	\includegraphics[trim={0.2cm 15.5cm 3cm 2.5cm},clip,scale=1]{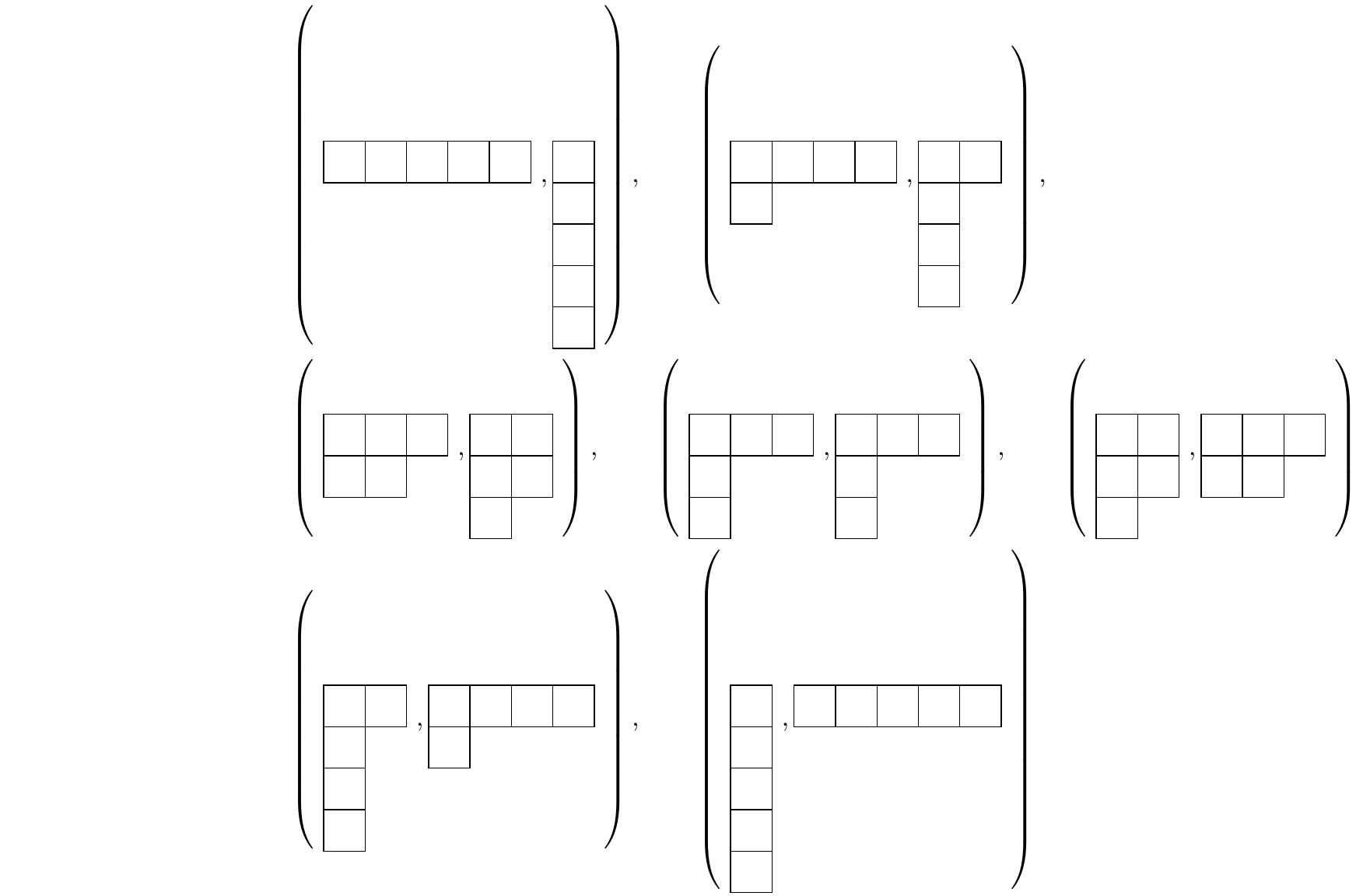}
\end{figure}\\
\begin{comment}
\bea
&&\left(\ \begin{ytableau}
	~&~&~&~&~
\end{ytableau}\ , \begin{ytableau}
~\\~\\~\\~\\~
\end{ytableau} \ \right),\qquad
\left(\ \begin{ytableau}
	~&~&~&~\\~
\end{ytableau}\ , \begin{ytableau}
~&~\\~\\~\\~
\end{ytableau} \ \right),\non\\
&& \left(\ \begin{ytableau}
	~&~&~\\~&~
\end{ytableau}\ , \begin{ytableau}
~&~\\~&~\\~
\end{ytableau} \ \right),\qquad  \left(\ \begin{ytableau}
	~&~&~\\~\\~
\end{ytableau}\ , \begin{ytableau}
	~&~&~\\~\\~
\end{ytableau} \ \right),\qquad
\left(\ \begin{ytableau}
	~&~\\~&~\\~
\end{ytableau}\ , \begin{ytableau}
~&~&~\\~&~
\end{ytableau} \ \right),\non\\
&& \left(\ \begin{ytableau}
	~&~\\~\\~\\~
\end{ytableau}\ , \begin{ytableau}
~&~&~&~\\~
\end{ytableau} \ \right),\qquad \left(\ \begin{ytableau}
~\\~\\~\\~\\~
\end{ytableau}\ , \begin{ytableau}
~&~&~&~&~
\end{ytableau} \ \right)
\eea
\end{comment}
\section{Three Slots up to Four Levels}
%\subsection{$g=(12)$}

In this Appendix, we list the fully anti-symmetric states for the case with three quantum numbers up to level 4. In level 4, we only show a sample state for brevity. 

At level 2, we have four different antisymmetric states:
\begin{figure}[h!]
	\centering
	\includegraphics[trim={4cm 24.2cm 3cm 2.5cm},clip,scale=1]{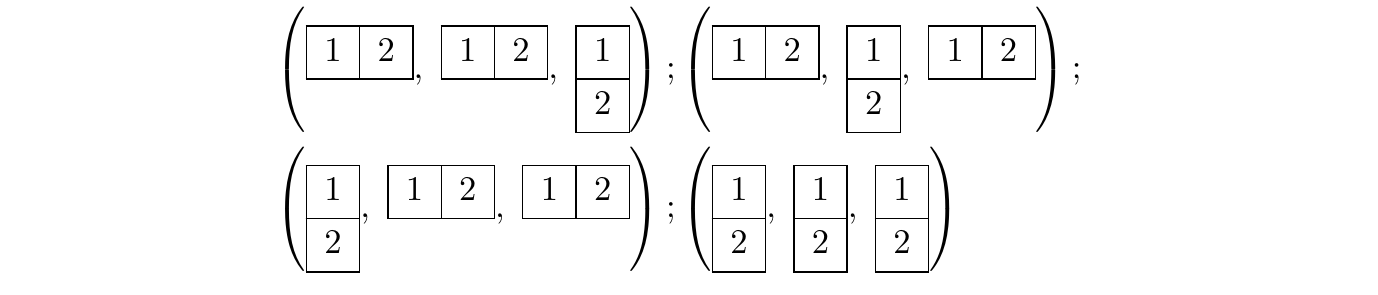}
\end{figure}\\
\begin{comment}
\begin{align}
\label{3slot-level2}
&\left(\begin{ytableau}
1&2\end{ytableau},~\begin{ytableau}
1&2 \end{ytableau},~\begin{ytableau}
1\\2 \end{ytableau}\right);
\left(\begin{ytableau}
1&2\end{ytableau},~\begin{ytableau}
1\\2 \end{ytableau},~\begin{ytableau}
1&2 \end{ytableau}\right); \non\\
&\left(\begin{ytableau}
1\\2\end{ytableau},~\begin{ytableau}
1&2 \end{ytableau},~\begin{ytableau}
1&2 \end{ytableau}\right);
\left(\begin{ytableau}
1\\2\end{ytableau},~\begin{ytableau}
1\\2 \end{ytableau},~\begin{ytableau}
1\\2 \end{ytableau}\right)
\end{align}
\end{comment}
%Equivalently, this can be thought of as a way of filling in 1 and 2 in an arbitrary 3-slot tableaux so that it is a part of an antisymmetric state. 
%\subsection{$g=(23)$}
At level 3, we have the following antisymmetric states:
\begin{figure}[h!]
	\centering
	\includegraphics[trim={4cm 18cm 3cm 2.5cm},clip,scale=1]{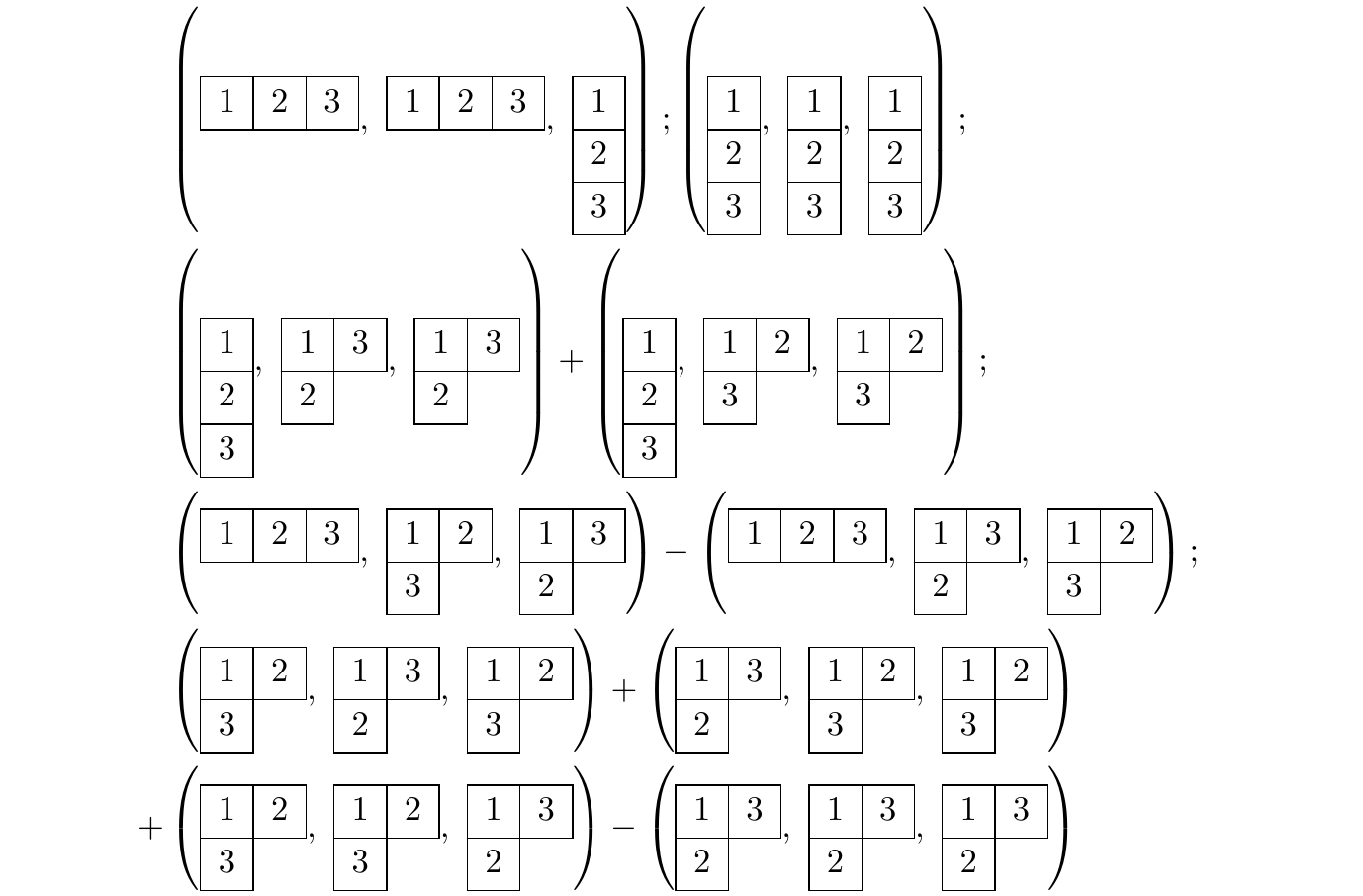}
\end{figure}\\
\begin{comment}
\begin{align}
&\left(\begin{ytableau}
1&2&3\end{ytableau},~\begin{ytableau}
1&2&3 \end{ytableau},~\begin{ytableau}
1\\2 \\3\end{ytableau}\right);
\left(\begin{ytableau}
1\\2\\3\end{ytableau},~\begin{ytableau}
1\\2\\3 \end{ytableau},~\begin{ytableau}
1\\2 \\3\end{ytableau}\right);\nonumber \\
&\left(\begin{ytableau}
1\\2\\3\end{ytableau},~\begin{ytableau}
1&3\\2 \end{ytableau},~\begin{ytableau}
1&3 \\2\end{ytableau}\right)+
\left(\begin{ytableau}
1\\2\\3\end{ytableau},~\begin{ytableau}
1&2\\3 \end{ytableau},~\begin{ytableau}
1&2 \\3\end{ytableau}\right);\nonumber \\
&\left(\begin{ytableau}
1&2&3\end{ytableau},~\begin{ytableau}
1&2\\3 \end{ytableau},~\begin{ytableau}
1&3 \\2\end{ytableau}\right)-
\left(\begin{ytableau}
1&2&3\end{ytableau},~\begin{ytableau}
1&3\\2 \end{ytableau},~\begin{ytableau}
1&2 \\3\end{ytableau}\right);\nonumber \\
&\left(\begin{ytableau}
1&2\\3\end{ytableau},~\begin{ytableau}
1&3\\2 \end{ytableau},~\begin{ytableau}
1&2 \\3\end{ytableau}\right)+
\left(\begin{ytableau}
1&3\\2\end{ytableau},~\begin{ytableau}
1&2\\3 \end{ytableau},~\begin{ytableau}
1&2 \\3\end{ytableau}\right)\nonumber \\
+&\left(\begin{ytableau}
1&2\\3\end{ytableau},~\begin{ytableau}
1&2\\3 \end{ytableau},~\begin{ytableau}
1&3 \\2\end{ytableau}\right)-
\left(\begin{ytableau}
1&3\\2\end{ytableau},~\begin{ytableau}
1&3\\2 \end{ytableau},~\begin{ytableau}
1&3 \\2\end{ytableau}\right)
\end{align}
\end{comment}
%\subsection{$g=(34)$}
\newpage
At level 4, the antisymmetric states are complicated and we will settle for showing just one of the states:
\begin{figure}[h!]
	\centering
	\includegraphics[trim={4cm 12cm 3cm 2.5cm},clip,scale=1]{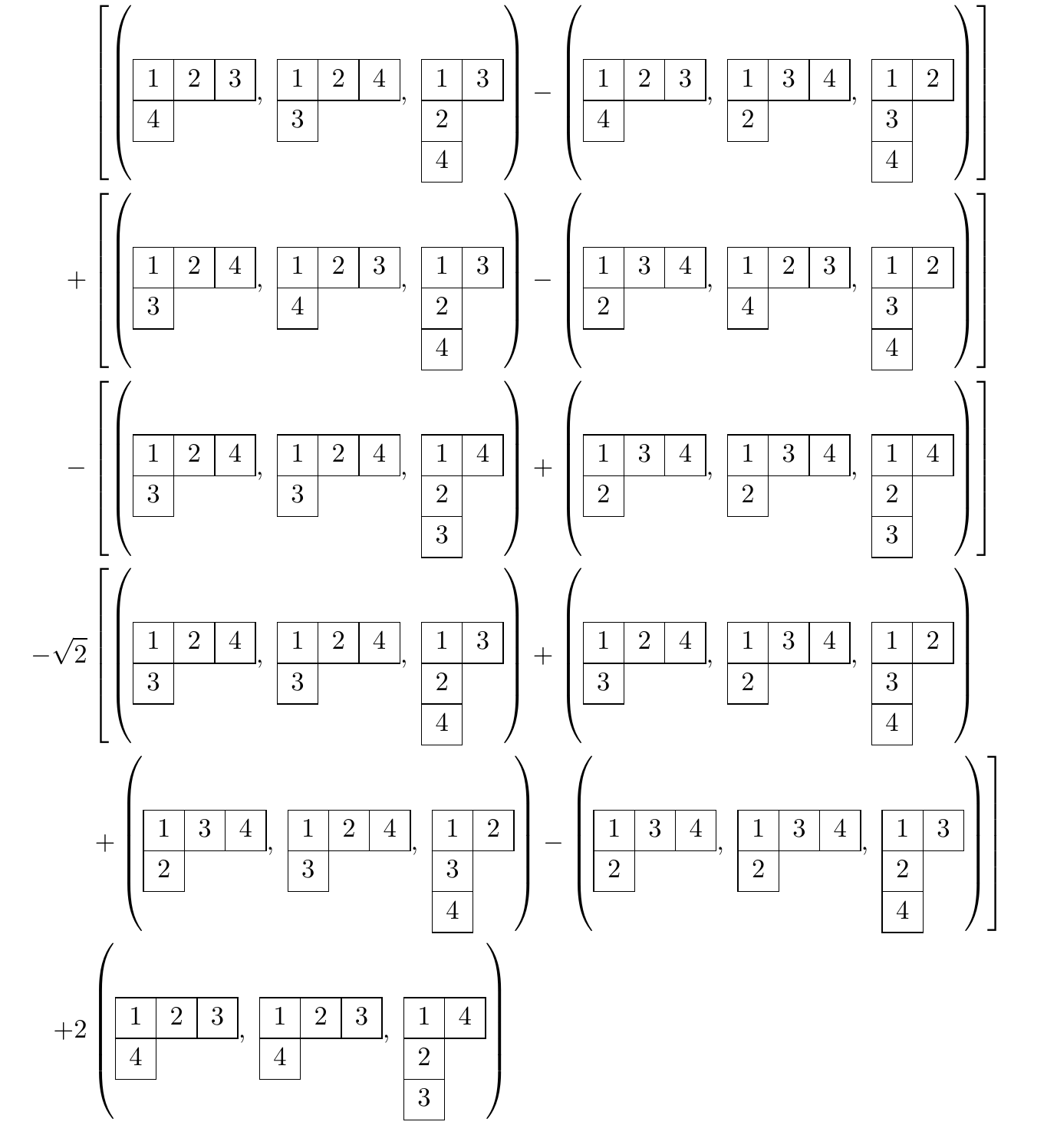}
\end{figure}\\
This state is an example where $(p_a,p_b,p_c)$ corresponding to $g_3=(34)$ are not constrained.

\section{Complete List of Anti-Symmetric Young Patterns at Level 4}\label{level4}

The multiplicities in front of the representations below are a short hand way of capturing the permutations of the Young pattern among the three slots. In particular, they are not meant to suggest actual multiplicities of the {\em same} representation.
\begin{figure}[h!]
	\centering
	\includegraphics[trim={4cm 5.2cm 3cm 2.5cm},clip,scale=1]{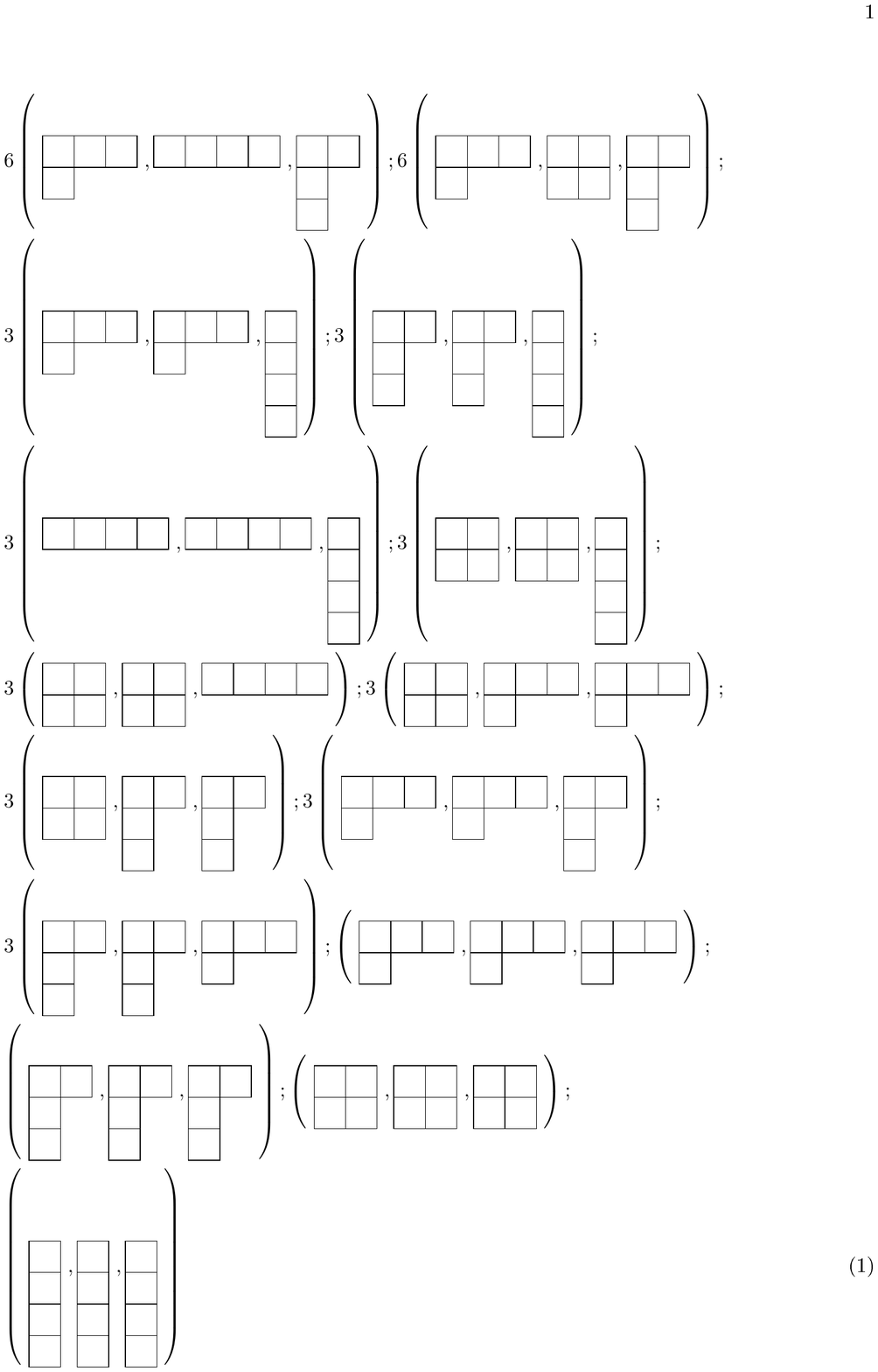}
\end{figure}\\

\newpage
\section{Complete List of Anti-Symmetric Young Patterns at Level 5}\label{level5}

As in level 4, (most of) the multiplicities in front of the representations below are a short hand way of capturing the permutations of the Young pattern among the three slots. In particular, (mostly) they are not meant to suggest actual multiplicities of the {\em same} representation.

But there are two exceptions to this. These are the two representations with bold face ${\bf 6}$'s: the number of permutations between the slots in each of those cases is 3. The extra factor of 2 actually denotes a multiplicity. This corresponds to the fact that there are two common eigenvectors that fall into those representations, in the language of section 3. 

\begin{figure}[htb!]
	\centering
	\includegraphics[trim={3cm 9.5cm 3cm 2.5cm},clip,scale=1]{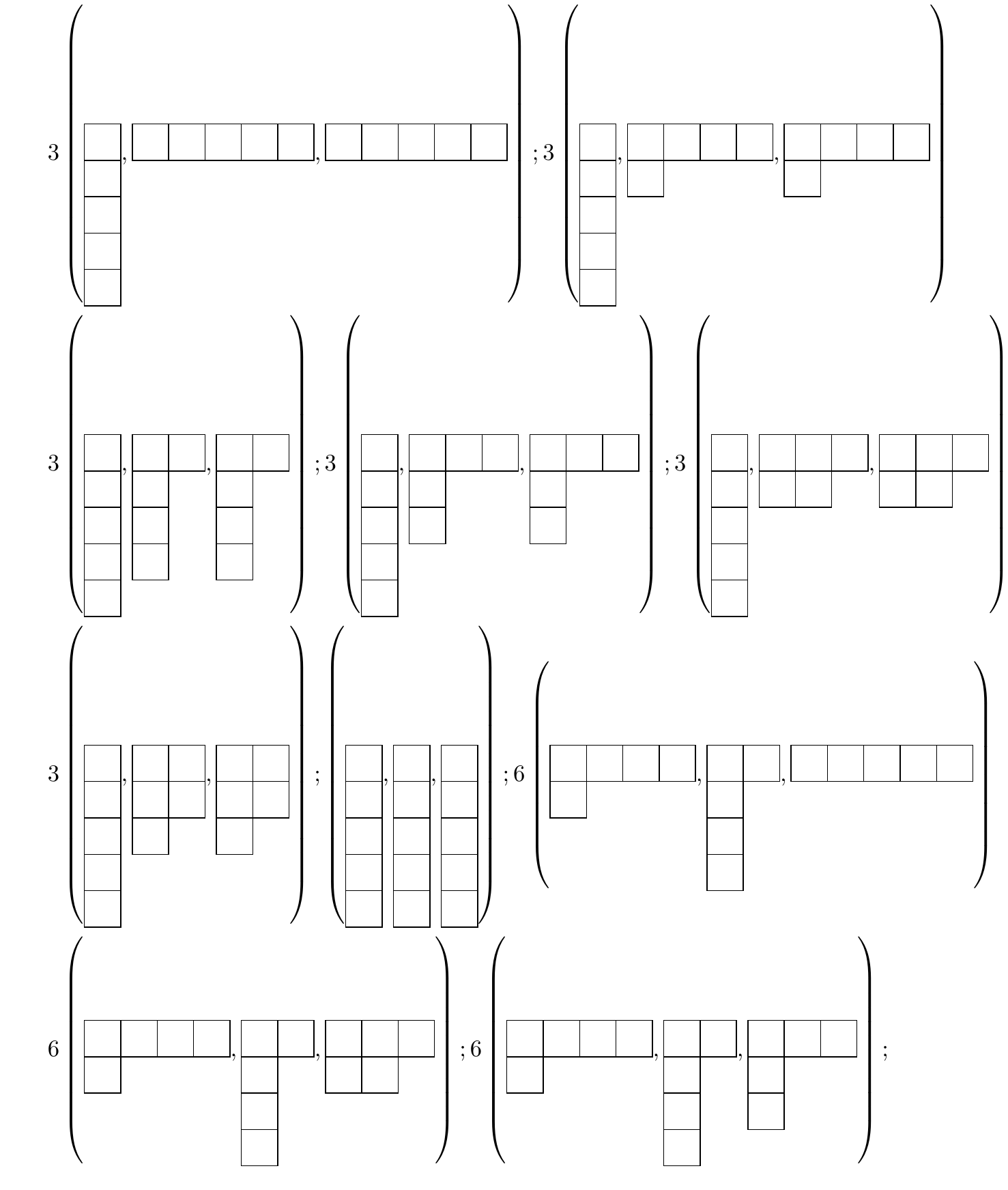}
\end{figure}
\begin{comment}
\begin{align}
3&\left(\begin{ytableau}~\\~\\~\\~\\~ \end{ytableau},\begin{ytableau}~&~&~&~&~ \end{ytableau},\begin{ytableau}~&~&~&~&~ \end{ytableau}\right);3\left(\begin{ytableau}~\\~\\~\\~\\~ \end{ytableau},\begin{ytableau}~&~&~&~\\~ \end{ytableau},\begin{ytableau}~&~&~&~\\~ \end{ytableau}\right)\nonumber \\
3&\left(\begin{ytableau}~\\~\\~\\~\\~ \end{ytableau},\begin{ytableau}~&~\\~\\~\\~ \end{ytableau},\begin{ytableau}~&~\\~\\~\\~ \end{ytableau}\right);3\left(\begin{ytableau}~\\~\\~\\~\\~ \end{ytableau},\begin{ytableau}~&~&~\\~\\~ \end{ytableau},\begin{ytableau}~&~&~\\~\\~ \end{ytableau}\right); 3\left(\begin{ytableau}~\\~\\~\\~\\~ \end{ytableau},\begin{ytableau}~&~&~\\~&~ \end{ytableau}, \begin{ytableau}~&~&~\\~&~ \end{ytableau}\right)\nonumber \\
3&\left(\begin{ytableau}~\\~\\~\\~\\~ \end{ytableau},\begin{ytableau}~&~\\~&~\\~ \end{ytableau}, \begin{ytableau}~&~\\~&~\\~ \end{ytableau}\right); \left(\begin{ytableau}~\\~\\~\\~\\~ \end{ytableau}, \begin{ytableau}~\\~\\~\\~\\~ \end{ytableau}, \begin{ytableau}~\\~\\~\\~\\~ \end{ytableau}\right);6\left(\begin{ytableau}~&~&~&~\\~ \end{ytableau}, \begin{ytableau}~&~\\~\\~\\~ \end{ytableau},\begin{ytableau}~&~&~&~&~ \end{ytableau}\right)\nonumber \\
6&\left(\begin{ytableau}~&~&~&~\\~ \end{ytableau}, \begin{ytableau}~&~\\~\\~\\~ \end{ytableau},\begin{ytableau}~&~&~\\~&~ \end{ytableau}\right);  6\left(\begin{ytableau}~&~&~&~\\~ \end{ytableau}, \begin{ytableau}~&~\\~\\~\\~ \end{ytableau},\begin{ytableau}~&~&~\\~\\~ \end{ytableau}\right);\nonumber 
\end{align}
\end{comment}

\begin{figure}[htbp!]
	\centering
	\includegraphics[trim={2cm 4.2cm 2.5cm 2.5cm},clip,scale=1]{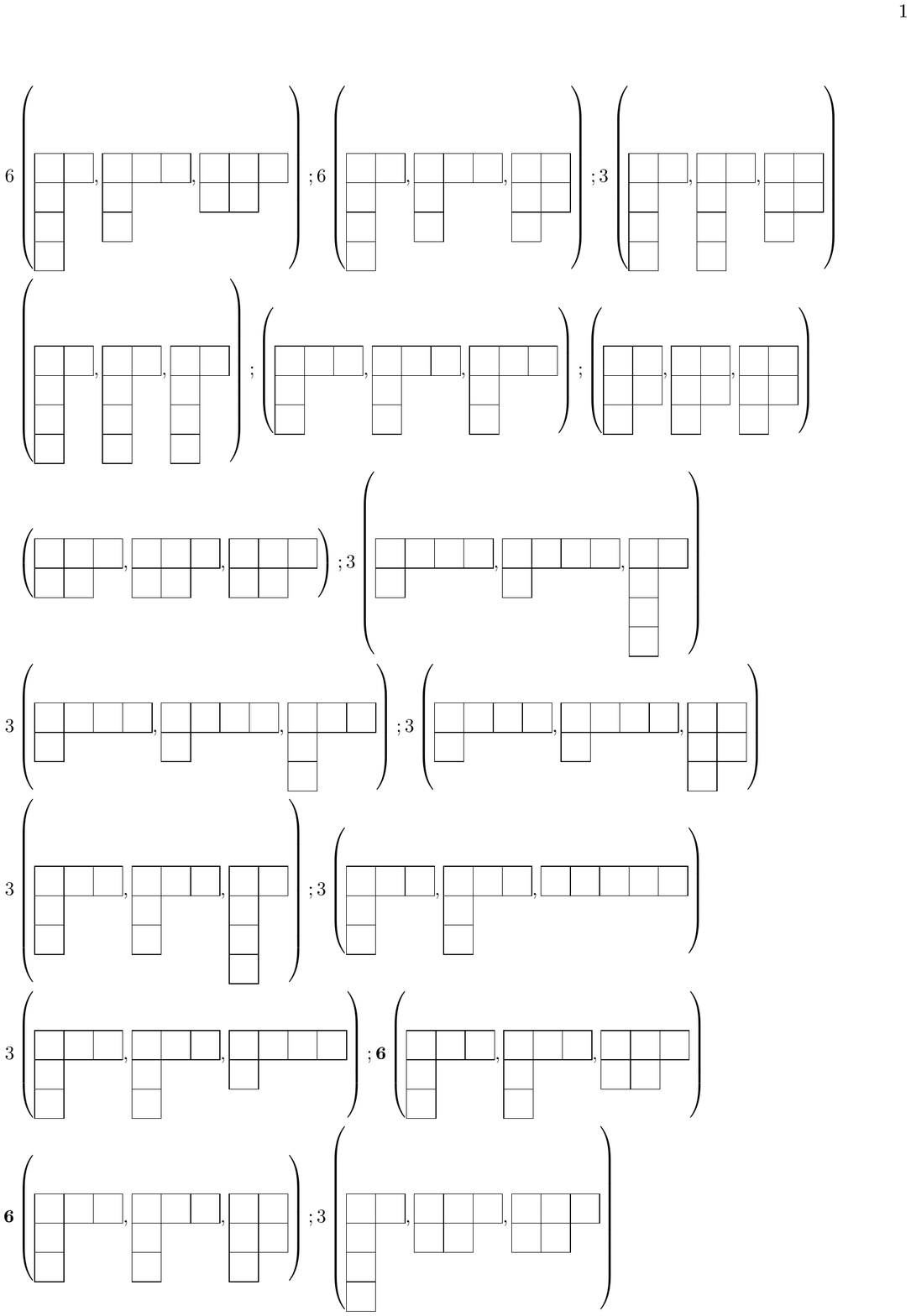}
\end{figure}

\begin{figure}[htbp!]
	\centering
	\includegraphics[trim={2cm 8.2cm 2.5cm 2.5cm},clip,scale=1]{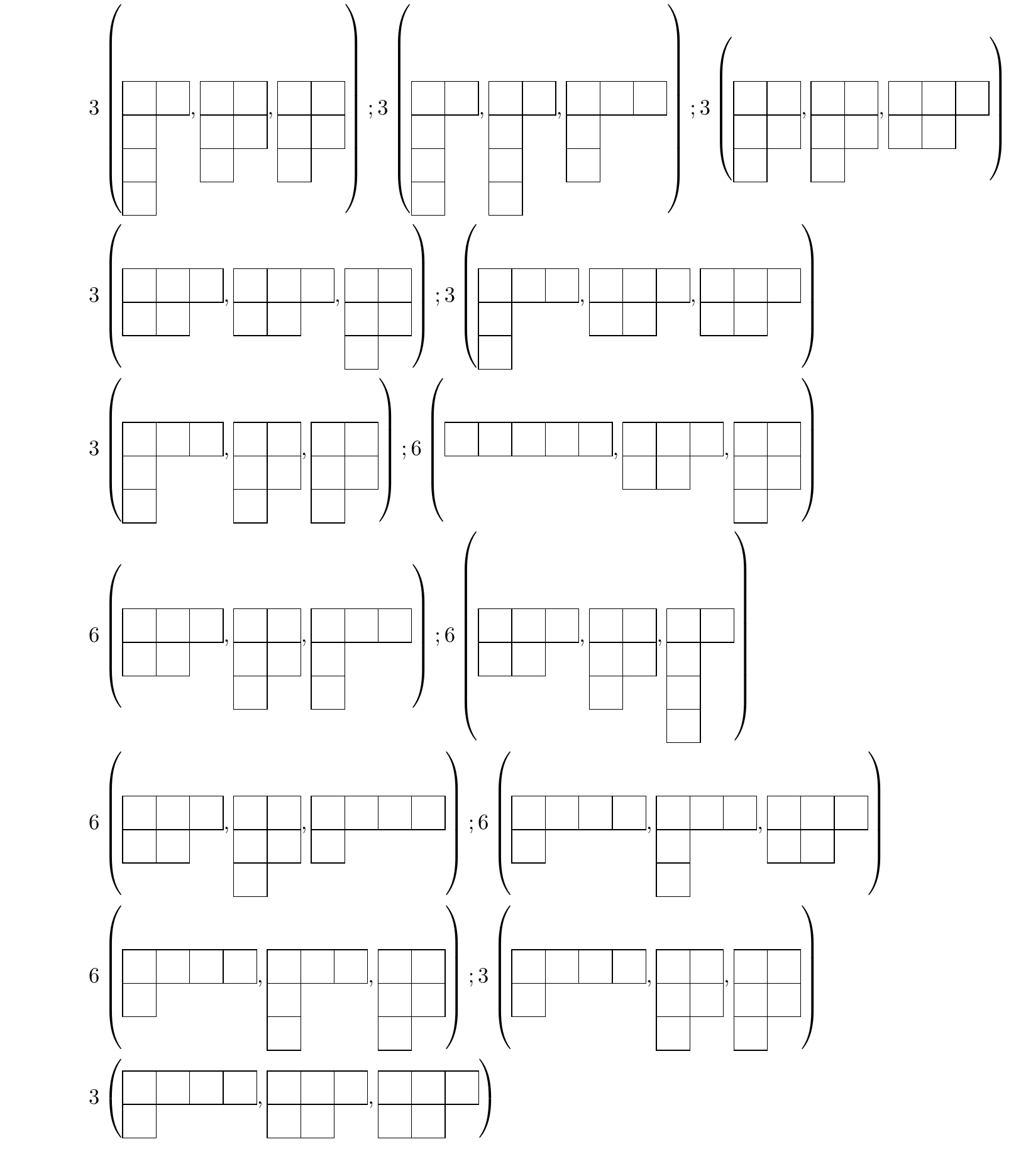}
\end{figure}

\end{document}